\documentclass[a4paper,11pt]{article}
\usepackage{bm,amssymb,amsmath,cite,color,verbatim,marvosym,graphicx,multicol,multirow}
\usepackage[colorlinks]{hyperref}

\def\lb{\label}
\def\ba{\begin{eqnarray}}
\def\ea{\end{eqnarray}}

\newcommand{\nc}{\newcommand}
\nc{\be}{\begin{equation}} \nc{\ee}{\end{equation}}
\nc{\bea}{\begin{eqnarray}} \nc{\eea}{\end{eqnarray}}
\nc{\disp}{\displaystyle} \nc{\ade}{\mbox{$A$-$D$-$E$}}
\nc{\calN}{{\cal N}} \nc{\calC}{{\cal C}} \nc{\calM}{{\cal M}}
\nc{\calS}{{\cal S}} \nc{\phit}{\hat{\varphi}}
\nc{\chit}{\hat{\chi}} \nc{\hcalN}{\hat{\calN}}
\nc{\hcalS}{\hat{\calS}} \nc{\hS}{\hat{S}}
\nc{\sigmad}{\sigma^\dagger} \nc{\psid}{\psi^\dagger}

\def\non{\nonumber}
\def\ncirc#1{\mbox{\textcircled{\tiny #1}}}

\nc{\todo}[1]{\textbf{( #1 )}}

\definecolor{IndianRed}{rgb}{0.8,0.36,0.36}
\def\ir{\color{IndianRed}}
\definecolor{DarkGreen}{rgb}{0,0.5,0}

\font\tenmsb=msbm10\font\sevenmsb=msbm7 \font\fivemsb=msbm5
\newfam\msbfam
\textfont\msbfam=\tenmsb \scriptfont\msbfam=\sevenmsb
\scriptscriptfont\msbfam=\fivemsb

\def\bra#1{\langle #1|}
\def\ket#1{|#1\rangle}

\renewcommand\eqref[1]{(\ref{#1})}

\DeclareMathOperator{\e}{e}

\DeclareMathOperator{\sign}{sign}
\def\ii{{\,\rm i}}

\numberwithin{equation}{section}

\begin{document}
\hypersetup{%
    urlcolor=IndianRed, 
    linkcolor=DarkGreen, 
    citecolor=DarkGreen 
} 

\begin{titlepage}
\title{Exact spin quantum Hall current between boundaries of a lattice strip}

\author{Jan de Gier$^a$\footnote{jdgier@unimelb.edu.au}, Bernard Nienhuis$^b$\footnote{b.nienhuis@uva.nl} and Anita Ponsaing$^a$\footnote{a.ponsaing@ms.unimelb.edu.au}
\bigskip\\
{\small\em
\begin{minipage}{\textwidth}
\begin{itemize}
\addtolength{\itemsep}{-\baselineskip}
\addtolength{\itemsep}{-\parskip}
\item[${}^a$] Department of Mathematics and Statistics, The University of Melbourne, VIC 3010, Australia\\
\item[${}^b$] Institute for Theoretical Physics, University of Amsterdam, The Netherlands
\end{itemize}
\end{minipage}}}

\maketitle

\begin{abstract}
Employing an inhomogeneous solvable lattice model, we derive an exact expression for a boundary-to-boundary edge current on a lattice of finite width. This current is an example of a class of parafermionic observables recently introduced in an attempt to rigorously prove conformal invariance of the scaling limit of critical two-dimensional lattice models. It also corresponds to the spin current at the spin-Quantum Hall transition in a model introduced by Chalker and Coddington, and generalized by Gruzberg, Ludwig and Read. Our result is derived from a solution of the $q$-deformed Knizhnik-Zamolodchikov equation, and is expressed in terms of a symplectic Toda-lattice wave-function.
\end{abstract}
{\footnotesize\tableofcontents}
\end{titlepage}

\section{Introduction}

In an attempt to rigorously prove conformal invariance of the scaling limit of critical two-dimensional lattice models, an interesting new class of parafermionic observables for Potts and loop models was recently introduced, see e.g. \cite{Smir06,Smir07,RivaCar,IkhCar}. These observables are expressed in terms of the loop representation of the Fortuin-Kasteleyn (FK) cluster expansion of the Potts-model \cite{BaxterKW76,Baxter}, and can be shown to be discretely holomorphic for certain parameter values. The operators corresponding to these observables carry a spin conjugate to the winding angles of the loops. In this paper we will compute an exact closed form expression for one such type of observable, with spin one, for a lattice loop model on a strip of finite width. As we show below, this observable corresponds to the spin-current at the quantum Hall transition in a system where time-reversal symmetry is broken, but $SU(2)$ spin symmetry is intact \cite{Gruzberg99}. The remarkable aspect of our result is the feasibility of an exact calculation for a finite size model which is not free-fermionic.

\begin{figure}[b!]
\centerline{\includegraphics[width=0.4\textwidth]{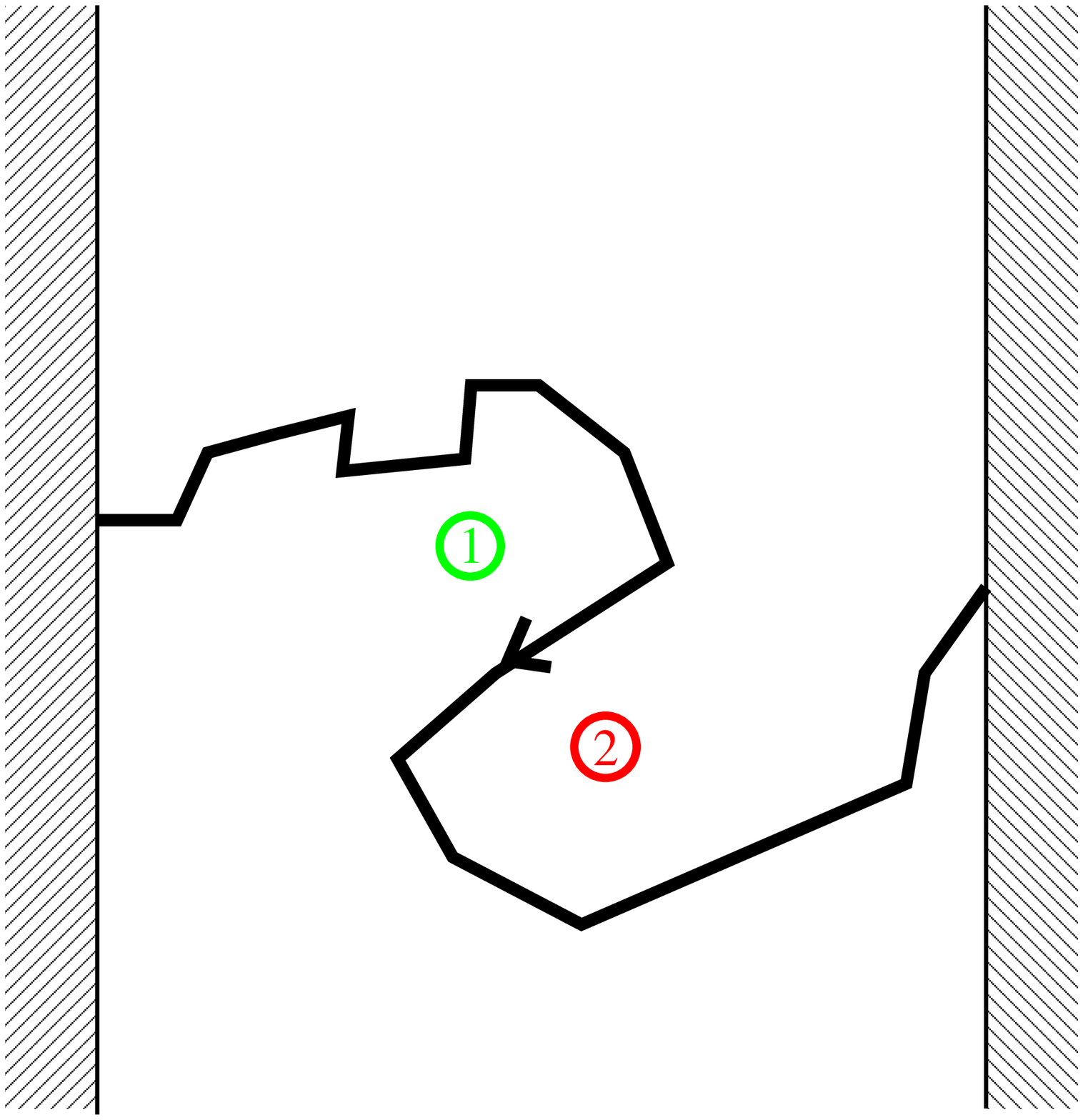}}
\caption{A path dividing the strip into two regions.}
\label{crack1}
\end{figure}
We will study a statistical ensemble of non-intersecting paths in a strip of finite width and infinite length. The paths may form closed loops, or terminate on the boundary. We assume that the paths that connect the two boundaries all carry an equal current from the left to the right boundary.  The closed loops, and the paths that connect two points of the same boundary do not carry any current. We wish to calculate the mean current density induced by the statistics of the paths. This can be expressed in the observable  $F^{(x_1,x_2)}$ representing the mean total current passing in between the points $x_1$ and $x_2$, oriented with $x_1$ to the left of the direction of the current.
\be
F^{(x_1,x_2)} = \sum_{C\in\Gamma} P(C) N^{(x_1,x_2)}_C \sign^{(x_1,x_2)}_C.
\label{flow_gendef}
\ee
Here $\Gamma$ is the set of configurations, $N_C^{(x_1,x_2)}$ is the number of paths passing in between the points $x_1$ and $x_2$ and running from the left to the right boundary, $P(C)$ is the ensemble probability of the configuration $C$ and $\sign_C^{(x_1,x_2)}$ is $+1$ if $x_1$ lies in the region above the paths, and $-1$ if it lies below. The sign of the path in Figure~\ref{crack1} thus is $-1$. The observable $F$ is antisymmetric, $F^{(x_1,x_2)}=-F^{(x_2,x_1)}$ and additive, $F^{(x_1,x_2)} + F^{(x_2,x_3)} = F^{(x_1,x_3)}$. 

The current defined above corresponds to the spin current in a generalised Chalker-Coddington network model \cite{ChalkC88,Metz99,Gruzberg99} for the spin quantum Hall effect. The Chalker-Coddington network model is based on the semi-classical picture of electrons in two dimensions moving under the influence of a strong perpendicular magnetic field in a long-ranged disorder potential. The network consists of a square lattice whose edges are unidirectional channels and whose vertices are scattering centers. A potential correlation length much larger than the magnetic length leads to the formation of clusters where the wave function amplitude changes only slightly. These clusters correspond to those of critical percolation. The classical limit of the generalised Chalker-Coddington model \cite{Metz99} on the square lattice is therefore described by the solvable completely packed O($n=1$) lattice model. In a further generalization Gruzberg et al \cite{Gruzberg99} used a pseudo spin description of the particle and hole states.  When the $SU(2)$ spin symmetry is observed, the full quantum mechanical spin current corresponds to the current observable we study here.  For these models we are able to derive a closed form expression for the ensemble expectation value of a discretised version of \eqref{flow_gendef}. The analogous calculation for the current around a cylinder has also been performed. \cite{Nienhuis10}

\section{The square lattice loop model for percolation}
Consider a square lattice of width $L$ and infinite height. Each square of the lattice has paths drawn on it in two possible orientations,
\[\raisebox{-13pt}{\includegraphics[height=30pt]{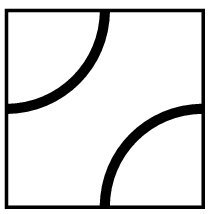}}\qquad\text{and}\qquad\raisebox{-13pt}{\includegraphics[height=30pt]{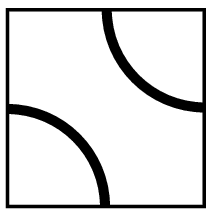}}\ ,\]
to which we assign probabilities $1/2$. These paths represent hulls of percolation clusters \cite{BaxterKW76}. In Section~\ref{weights} we will consider much more general relative probabilities, but we postpone introducing technicalities for clarity of exposition. We also assign probabilities to the paths at the edge of the lattice, which can be connected either to the boundary or to each other. Paths at the left boundary correspond to the following pictures,  
\[\raisebox{-25pt}{\includegraphics[height=60pt]{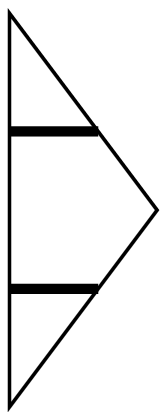}}\qquad\text{and}\qquad\raisebox{-25pt}{\includegraphics[height=60pt]{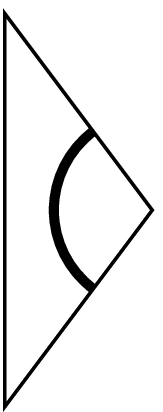}}\ ,\]
and likewise for the right boundary. In this way the lattice rows come in groups of two:
\[\raisebox{-40pt}{\includegraphics[height=70pt]{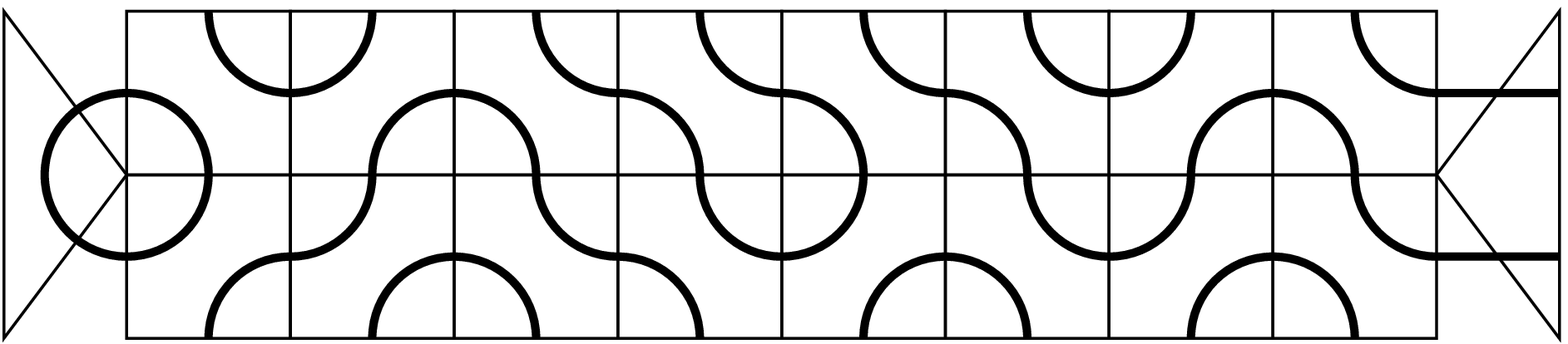}}.\]
Each site on the upper or lower edge of a double row is connected to either the boundary or another site. Paths with both ends connected to one boundary and those connecting one boundary to the other both have weight one. We denote the weight of closed loops in the bulk by $-(q+1/q)$, which introduces the parameter $q$. We will mainly be interested in the special point $q=\e^{2\pi \ii/3}$, and so closed loops will also have weight $1$.
The way in which paths connect the sites on either the top or bottom edge of a double row is called a link pattern. 

When we consider the model on the semi-infinite strip, e.g. the lattice extends infinitely far below, sites on the top edge are either connected to each other, or to one of the boundaries. The total number of such link patterns in a system of size $L$ is $2^L$. Each link pattern can be expressed using open and closing parentheses where matching parentheses represent connected sites, unmatched opening (closing) parentheses indicate a connection to the right (left) boundary. For example, $()()(()($ is the link pattern of size $L=8$ shown on the top edge above. This model is known as the completely packed O($n=1$) loop model.

\subsection{Boundary-to-boundary current}
We now define the boundary-to-boundary current for the completely packed O($n=1$) model. The ensemble of paths in this case is given by the possible configurations of the bulk and boundary faces of the lattice, and $P(C)$ in \eqref{flow_gendef} is uniform for all possible configurations. 
\begin{figure}[h]
\centerline{\includegraphics[height=150pt]{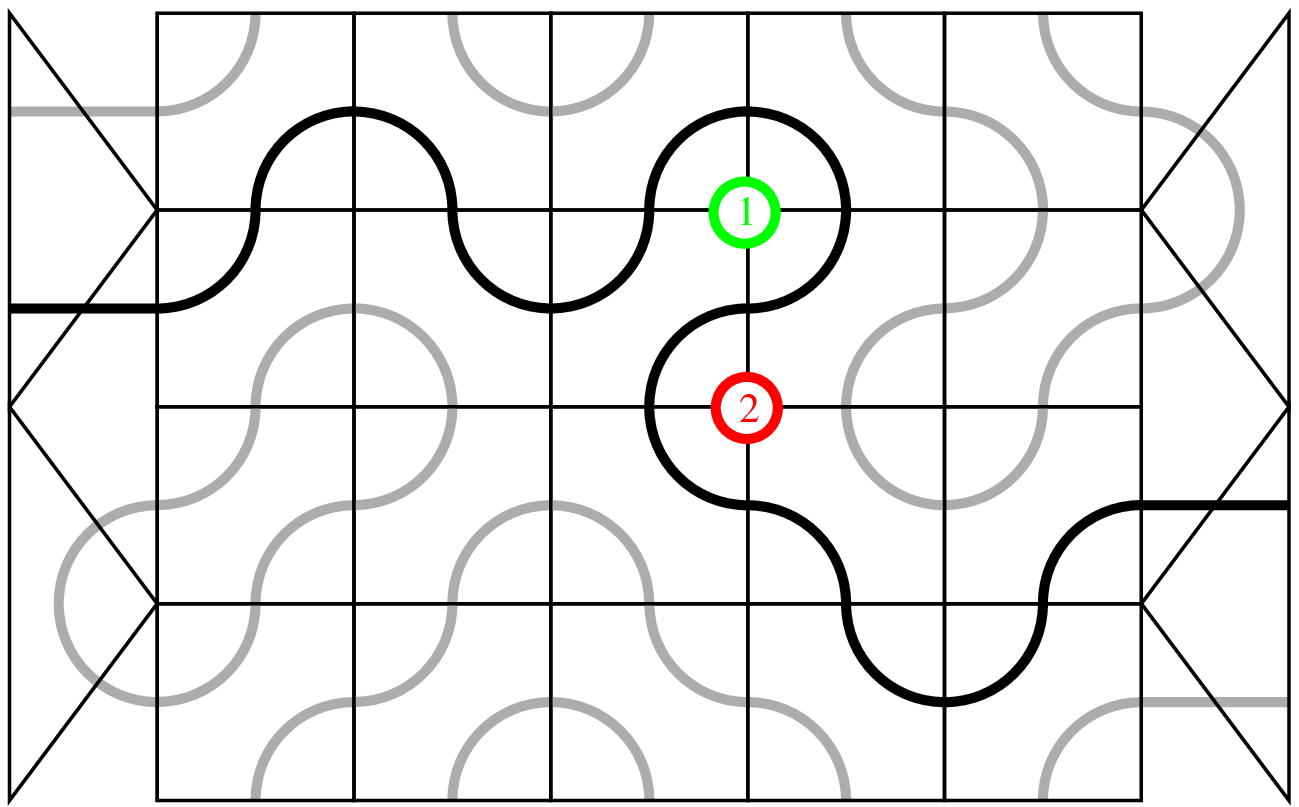}}
\caption{Example path between markers 1 and 2 with phase $-1$.}
\label{Vflow_O(1)}
\end{figure}

Because of the additivity property of $F$ it suffices to concentrate on the case where the two markers are placed on adjacent sites, so that  $|N_C^{(x_1,x_2)}| \leq 1$.  The markers can be separated by a horizontal lattice edge
\be
\label{eq:X}
X^{(k)} = F^{((i,k),(i,k+1))},
\ee
or a vertical one, as in Figure~\ref{Vflow_O(1)}
\be
\label{eq:Y}
Y^{(k)} = F^{((k,j),(k-1,j))}.
\ee
%

\subsection{Main result}
Our main result is an explicit expression for \eqref{eq:X} and \eqref{eq:Y} on an inhomogeneous lattice infinite in both vertical directions but of finite width. In order to present these expressions here we need to first introduce some auxiliary functions. Let
\be
\tau_L(z_1,\ldots,z_L) = \chi_{\lambda^{(L)}}(z_1^2,\ldots,z_L^2),\qquad \lambda^{(L)}_j = \left\lfloor \frac{L-j}{2}\right\rfloor, 
\label{Sdef}
\ee
where we have introduced the symplectic character $\chi_\lambda$ of degree $\lambda$ defined by
\be
\chi_\lambda(z_1,\ldots,z_L) = \frac{\left|z_i^{\lambda_j+L-j+1}-z_i^{-\lambda_j-L+j-1}\right|_{1\leq i,j\leq L}}{\left|z_i^{L-j+1}-z_i^{-L+j-1}\right|_{1\leq i,j\leq L}}.
\lb{sympchar}
\ee
Also let
\be
u_L(\zeta_1,\zeta_2;z_1,\ldots,z_L)=c_L \log \left[ \frac{\tau_{L+1}(\zeta_1,z_1,\ldots,z_L)\ \tau_{L+1}(\zeta_2,z_1,\ldots,z_L)} {\tau_{L}(z_1,\ldots,z_L)\ \tau_{L+2}(\zeta_1,\zeta_2,z_1,\ldots,z_L)} \right],
\label{wave}
\ee
where the constant $c_L$ is given by
\be
c_L=(-)^L \ii \frac{\sqrt{3}}{2}.
\ee
It is worthwhile to point out that the function $u_L$ defined in \eqref{wave} precisely has the form of a Toda lattice wave function \cite{JimboM}. Under a technical assumption explained in Section~\ref{degree}, we will prove that, on an inhomogeneous lattice of width $L$,
\begin{align}
X^{(k)}_L &= \ z_k 
\frac{\partial}{\partial z_k}\ u_L(\zeta_1,\zeta_2;z_1,\ldots,z_L)
,
\label{Xval}\\
Y^{(k)}_L & = 
\left.\ w \frac{\partial}{\partial w}\ u_{L+2}(\zeta_1,\zeta_2;z_1,\ldots,z_L,vq^{-1},w)\right|_{v=w}
\quad ({\rm independent\; of\;}k),
\label{Yval}
\end{align}
where $v$ is a dummy variable and $w$ is a spectral parameter introduced by the transfer matrix of the system. We will now describe in more details the dependence on the parameters $z_1,\ldots,z_L$ which are attached to the $L$ sites, and $\zeta_1,\zeta_2$ to the left and right boundary respectively.

\subsection{Inhomogeneous weights}
\label{weights}
In this section we define a much more general version of the O($n$) model. We define
\be
k(a,b)=[q/ab][qb/a], \qquad\qquad [z]=z-z^{-1},
\ee
and parametrise the relative probabilities of orientations of paths on elementary plaquettes by
\begin{align}
R(z,w)&=\frac{[qz/w]}{[qw/z]}\quad\raisebox{-18pt}{\includegraphics[height=40pt]{R1.eps}}+\;\frac{[z/w]}{[qw/z]}\quad\raisebox{-18pt}{\includegraphics[height=40pt]{R2.eps}}\non\\
&=:\ \raisebox{-37pt}{\includegraphics[height=80pt]{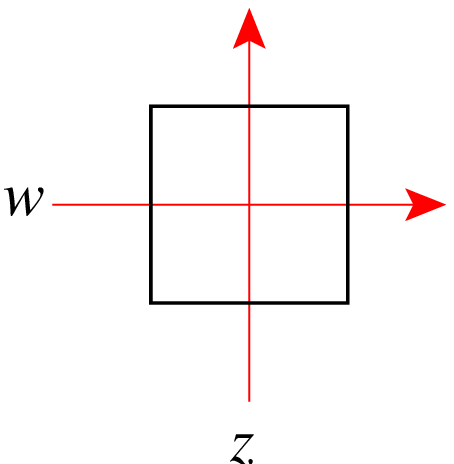}}.
\end{align}
Correspondingly, for path segments at the boundaries we take
\begin{align}
K_{\rm r}(w,\zeta)&=\frac{k(w,\zeta)}{k(1/w,\zeta)}\quad\raisebox{-28pt}{\includegraphics[height=60pt]{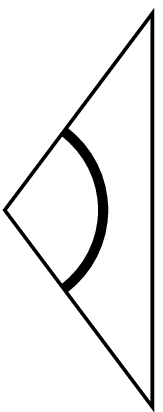}}\;-\frac{[q][w^2]}{k(1/w,\zeta)}\quad\raisebox{-28pt}{\includegraphics[height=60pt]{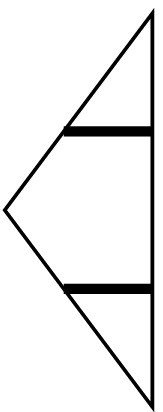}}\non\\
&=:\ \raisebox{-28pt}{\includegraphics[height=60pt]{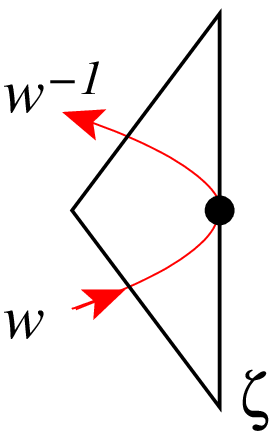}},\\
K_{\rm l}(w,\zeta)&=\frac{k(q/w,\zeta)}{k(w/q,\zeta)}\quad\raisebox{-28pt}{\includegraphics[height=60pt]{K01.eps}}\;-\frac{[q][q^2/w^2]}{k(w/q,\zeta)}\quad\raisebox{-28pt}{\includegraphics[height=60pt]{K02.eps}}\non\\
&=:\ \raisebox{-28pt}{\includegraphics[height=60pt]{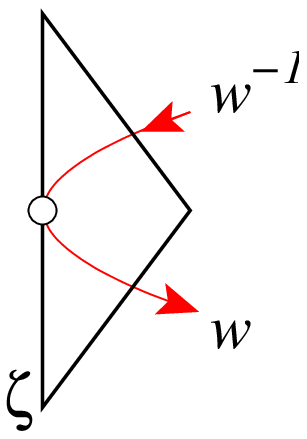}}.
\end{align}
Multiplication of two pictures consists of concatenation and identification of edges in such a way that lines with arrows join up according to orientation, see e.g. \cite{GierPS}. It then follows that these $R$ and $K$ matrices satisfy
\be
\label{eq:unit}
\begin{array}{rl}
R(z,w)R(w,z)&=1\\
K_{\rm r}(w,\zeta)K_{\rm r}(1/w,\zeta)&=1,\qquad K_{\rm l}(qw^{-1},\zeta)K_{\rm l}(qw,\zeta)=1.
\end{array}
\ee
Note that we can use these pictures in any orientation, as the arrows uniquely determine how the variables $z$ and $w$ enter. The transfer matrix for the generalised system is then
\[T_L(w) = \raisebox{-70pt}{\includegraphics[height=140pt]{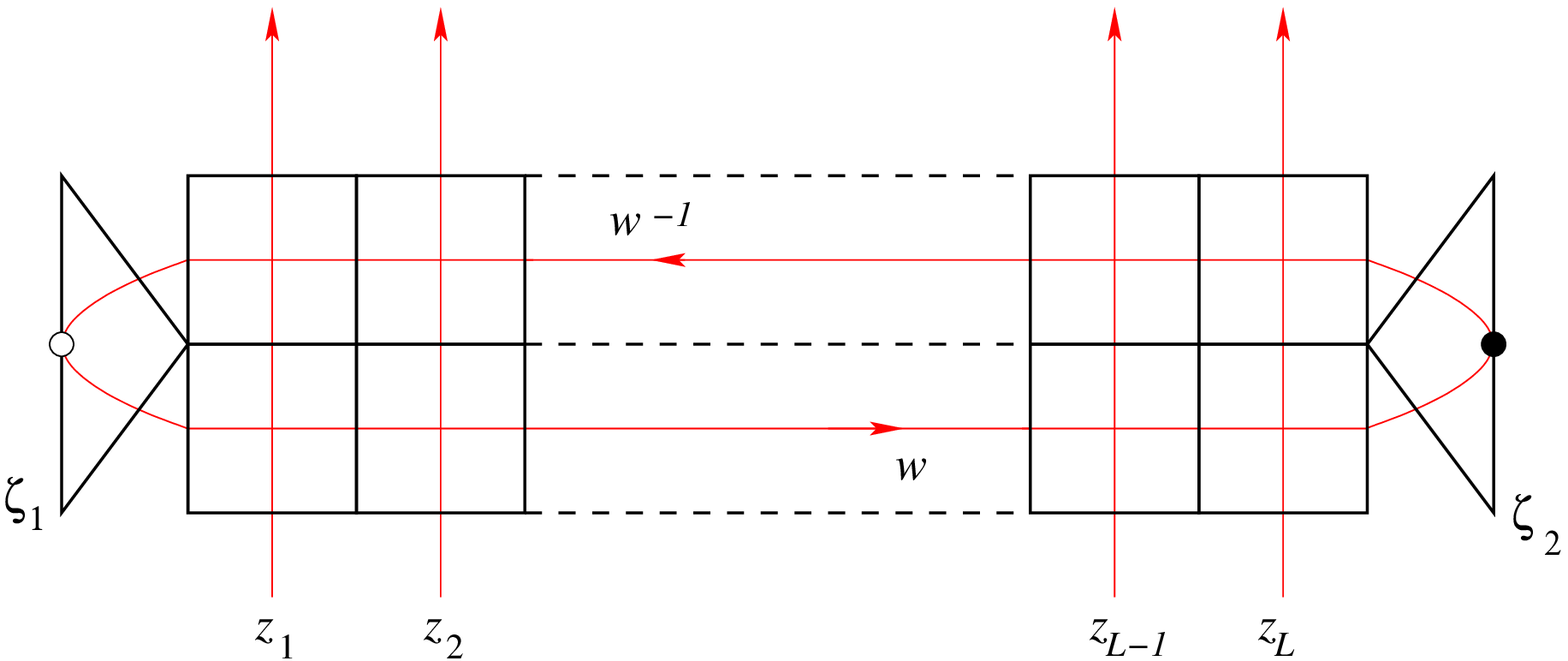}}\;.\]

In \cite{Skly88} it was shown that such double row transfer matrices form a one parameter family of commuting matrices, see also \cite{GierPS,RasmusP} in the context of loop models. The transfer matrix acts on a link pattern by sending it to a linear combination of other link patterns, weighted by (Laurent) polynomials in the parameters $z_i$ and $\zeta_j$ introduced by the $R$ and $K$ matrices. Note that the polynomials in the eigenvectors of the transfer matrix do not depend on $w$, which is a spectral parameter only.

\subsection{Infinite lattice and groundstate eigenfunction}
Consider a horizontal line across the infinite lattice of finite width. The probability distribution of downward link patterns on this line is given by the groundstate eigenvector of the transfer matrix, which for $q=\e^{2\ii\pi/3}$ has corresponding eigenvalue $1$ \cite{GierPS}. We will denote this groundstate eigenvector by $\ket{\Psi(\zeta_1,\zeta_2;z_1,\ldots,z_L)}$,
\be
T_L(w;\zeta_1,\zeta_2;z_1,\ldots,z_L)\ \ket{\Psi(\zeta_1,\zeta_2;z_1,\ldots,z_L)} = \ket{\Psi(\zeta_1,\zeta_2;z_1,\ldots,z_L)}.
\ee
This is a vector of polynomials in $z_i$ and $\zeta_j$, indexed by link patterns $\alpha$:
\be
\ket{\Psi(\zeta_1,\zeta_2;z_1,\ldots,z_L)}=\sum_\alpha \psi_\alpha(\zeta_1,\zeta_2;z_1,\ldots,z_L)\ket\alpha.\\
\ee

The probabilities for the upward link patterns are prescribed by the transpose vector $\bra{\Psi^*}$. Since we have
\be
\label{Ttranspose}
T_L(w;\zeta_1,\zeta_2;z_1,\ldots,z_L)^t = T_L(qw^{-1};\zeta_2,\zeta_1;z_L,\ldots,z_1),
\ee
the vector $\bra{\Psi^*}$ is related to $\ket\Psi$ by a horizontal reflection of the link patterns [eg., $)(()()$ goes to $()())($ for $L=6$] and a swapping of parameters; $\zeta_1\leftrightarrow\zeta_2$, $z_1\leftrightarrow z_L$, $z_2\leftrightarrow z_{L-1},\ldots$ . So we have
\be
\bra{\Psi^*(\zeta_1,\zeta_2;z_1,\ldots,z_L)} = \ket{\Psi(\zeta_2,\zeta_1;z_L,\ldots,z_1)}^t.
\ee
Note that the dual vector $\bra{\Psi^*}$ should not be confused with the left eigenvector of the transfer matrix, which for $q=\e^{2\ii\pi/3}$ is simply the vector whose entries are all equal. 

We will at times suppress some of the arguments of polynomials. Unless otherwise stated, $Y_L=Y_L(w;\zeta_1,\zeta_2;z_1,\ldots,z_L)$ with a special dependence on $w$, and $X_L^{(k)}=X_L^{(k)}(\zeta_1,\zeta_2;z_1,\ldots,z_L)$ with a special dependence on $z_k$. Where relevant, variables will be given explicitly, and the notation $\hat z_i$ in a list of arguments will mean that $z_i$ is missing from the list. 

Let us define the operator $\widehat{F}_L^{(x_1,x_2)}$, where $x_1$ and $x_2$ are separated by one horizontal or vertical step, by marking two vertices inside the transfer matrix $T_L$. For example, $\widehat{Y}^{(3)}_L$ is depicted as
\be
\widehat{Y}_L^{(3)} = \raisebox{-70pt}{\includegraphics[height=140pt]{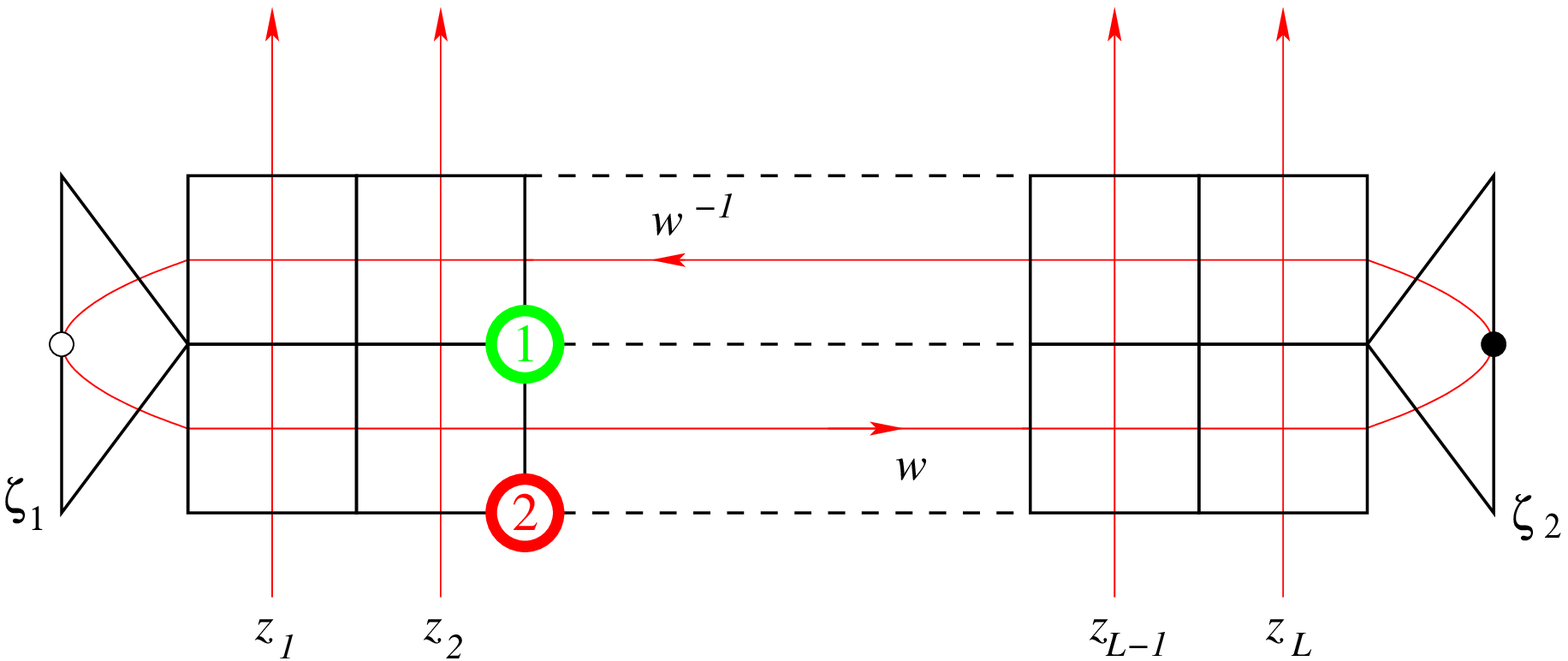}}\;.
\ee
The expectation value $F_L^{(x_1,x_2)}$, as defined in \eqref{flow_gendef}, is thus given for the infinite lattice by
\be
\label{Expectation}
F_L^{(x_1,x_2)}=\frac{\bra{\Psi^*}\ \widehat{F}_L^{(x_1,x_2)}\ \ket{\Psi}}{\bra{\Psi^*}\Psi\rangle }.
\ee
As $\bra{\alpha}\beta\rangle=1$, $\forall \alpha,\beta$, we have that
\be
\bra{\Psi^*}\Psi\rangle = \sum_{\alpha,\beta}\psi^*_{\alpha}\psi_{\beta} = Z_L^*Z_L =Z_L^2,
\ee
where $Z_L=\sum_\alpha \psi_\alpha$ is the normalisation of the eigenvector. It can be shown, see \cite{GierPS} as well as the next section, $Z_L^*=Z_L$. 

A special case of this operator is when both markers are on the top or bottom row of the transfer matrix (these are equivalent). In this case the coordinates of the markers are of the form $x_1=(k,j)$ and $x_2=(k+1,j)$, and we can write the operator as a product. For example, when the markers sit at the bottom row of a transfer matrix, we can write $\widehat{F}^{(x_1,x_2)}_L = \widehat{X}^{(k)}_L = T_L^t\ \kappa^{(k)}_X$, where $\kappa$ is the operator consisting of the two markers. Hence we have
\be
X^{(k)}_L =\frac1{Z_L^2}\bra{\Psi^*} \widehat{X}^{(k)}_L\ket\Psi
=\frac1{Z_L^2}\bra{\Psi^*}T_L^t\ \kappa^{(k)}_X \ket\Psi
=\frac1{Z_L^2}\bra{\Psi^*}\kappa^{(k)}_X\ket\Psi.
\ee
The expectation value $X^{(k,{\rm bot})}_L$ therefore does not depend on the spectral parameter $w$.

The pair of markers in $\widehat{X}$ has one other possible vertical position in the double row transfer matrix -- on the middle line. We will show that this case is equal to markers at the bottom. There are two possible vertical positions for the markers in $\widehat{Y}$, but the top case is obtained from the bottom case by sending $z_i\rightarrow 1/z_i$. We therefore concentrate on the operators $\widehat{X}$ and $\widehat{Y}$ where the positions of the markers can be expressed purely in terms of their horizontal position. The horizontal position of the \ncirc{1} marker is given by $k$, and the \ncirc{2} marker is either to the right of \ncirc{1} in the case of $\widehat{X}$, or below in the case of $\widehat{Y}$. We define the positions of the vertices of the lattice by assigning $1$ to $L+1$ horizontally from left to right. In this way the position $k=2$ refers to the point one column along the bottom row of the transfer matrix. This numbering convention has the advantage that the parameter whose rapidity line passes between the markers in $\widehat X^{(k)}$ is $z_k$.

In the next section we will also show that $Y_L$ is independent of the horizontal position of $\widehat Y$, so we will only use the $k$ label for $\widehat X$. 

%
\section{Symmetries}
\label{symmetries}
The expectation values $Y_L$ and $X_L^{(i)}$ have a number of symmetries, most arising from properties of the transfer matrix. The transfer matrix also allows us to deduce certain recursion relations which the expectation values should satisfy, and the symmetries are then used to generalise these relations. In this section we describe all the symmetries and relations satisfied by the expectation values.

\subsection{Symmetries of the expectation values}
\label{sec:symm}
As shown in \cite{DF05,GierPS}, and suppressing irrelevant notation, the transfer matrix satisfies the interlacing conditions 
\begin{align}
R(z_i,z_{i+1})T_L(\ldots,z_i,z_{i+1},\ldots) &= T_L(\ldots,z_{i+1},z_i,\ldots)R(z_i,z_{i+1}),\non\\
K_{\rm l}(qz_1,\zeta_1)T_L(z_1,\ldots) &= T_L(z_1^{-1}\ldots)K_{\rm l}(qz_1,\zeta_1),
\label{Tinterlace}\\
K_{\rm r}(z_L,\zeta_2)T_L(\ldots,z_L) &= T_L(\ldots,z_L^{-1})K_{\rm r}(z_L,\zeta_2).\non
\end{align}
It is easily seen that these interlacing conditions still hold to the left and right of the markers in the operators $\widehat X^{(k)}$ or $\widehat Y^{(k)}$, for example
\be
R(z_i,z_{i+1})\widehat{X}^{(k)}_L(\ldots,z_i,z_{i+1},\ldots) = \widehat{X}^{(k)}_L(\ldots,z_{i+1},z_i,\ldots)R(z_i,z_{i+1}),
\ee
for all values $k\neq i,i+1$.

Equations \eqref{Tinterlace} and the uniqueness of the groundstate eigenvector $\ket\Psi$  lead to following equations which constitute the $q$-Knizhnik-Zamolodchikov equation \cite{smirnov,FR} for open boundaries \cite{DF05,ZJ07,GierPS,Cant},
\begin{align}
R(z_i,z_{i+1})\ket{\Psi(\ldots,z_i,z_{i+1},\ldots)}&=\ket{\Psi(\ldots,z_{i+1},z_i,\ldots)}\non\\
K_{\rm l}(qz_1,\zeta_1)\ket{\Psi(z_1,\ldots)}&=\ket{\Psi(z_1^{-1},\ldots)}
\label{qKZ}\\
K_{\rm r}(z_L,\zeta_2)\ket{\Psi(\ldots,z_L)}&=\ket{\Psi(\ldots,z_L^{-1})}.\non
\end{align}
Similarly, the dual vector $\bra{\Psi^*}$ satisfies
\begin{align}
\bra{\Psi^*(\ldots,z_i,z_{i+1},\ldots)} R(z_{i+1},z_i)&=\bra{\Psi^*(\ldots,z_{i+1},z_i,\ldots)}\non\\
\bra{\Psi^*(z_1,\ldots)}K_{\rm l}(qz_1^{-1},\zeta_1)&=\bra{\Psi^*(z_1^{-1},\ldots)}
\label{qKZ*}\\
\bra{\Psi^*(\ldots,z_L)}K_{\rm r}(z_L^{-1},\zeta_2)&=\bra{\Psi^*(\ldots,z_L^{-1})}.\non
\end{align}

Using \eqref{eq:unit}, we thus have for $k\neq i,i+1$,
\begin{align}
X_L^{(k)}(\ldots,z_i,z_{i+1},\ldots) = \bra{\Psi^*} \widehat{X}^{(k)}_L\ket\Psi &=\bra{\Psi^*} R(z_{i+1},z_i)R(z_i,z_{i+1})\widehat{X}^{(k)}_L\ket\Psi\non\\
&=\bra{\Psi^*(z_{i+1},z_i)}\widehat{X}^{(k)}_L(z_{i+1},z_i)R(z_i,z_{i+1})\ket\Psi\non\\
&=\bra{\Psi^*(z_{i+1},z_i)}\widehat{X}^{(k)}_L(z_{i+1},z_i)\ket{\Psi(z_{i+1},z_i)}\non\\
&= X_L^{(k)}(\ldots,z_{i+1},z_{i},\ldots).
\end{align}
and similarly, for $k\neq 1$ and $k\neq L$ respectively,
\be
\renewcommand{\arraystretch}{1.6}
\begin{array}{rl}
X_L^{(k)}(z_1,\ldots) = \bra{\Psi^*} \widehat{X}^{(k)}_L\ket\Psi  &= \bra{\Psi^*} K_{\rm l}(qz_1^{-1},\zeta_1)K_{\rm l}(qz_1,\zeta_1)\widehat{X}^{(k)}_L\ket\Psi \\
&= \bra{\Psi^*} K_{\rm r}^t(z_1,\zeta_1)\widehat{X}^{(k)}_L(z_1^{-1},\ldots) K_{\rm l}(qz_1,\zeta_1)\ket\Psi \\
&= \bra{\Psi^*(z_1^{-1},\ldots,)}\widehat{X}^{(k)}_L(z_1^{-1},\ldots)\ket{\Psi(z_1^{-1},\ldots,)}\\
&= X_L^{(k)}(z_1^{-1},\ldots),\\
X_L^{(k)}(\ldots,z_L) = \bra{\Psi^*} \widehat{X}^{(k)}_L\ket\Psi &= \bra{\Psi^*(\ldots,z_L^{-1})}\widehat{X}^{(k)}_L(\ldots,,z_L^{-1})\ket{\Psi(\ldots,z_L^{-1})}\\
&=X_L^{(k)}(\ldots,z_L^{-1}).
\end{array}
\ee
Therefore, $X^{(k)}_L\in\mathbb{C}[z_1^{\pm},\ldots,z_{k-1}^{\pm}]^{W_{\rm B}}$, i.e., regarding the variables $z_k,\ldots,z_L$ as complex numbers, $X^{(k)}_L$ is invariant under the action of the Weyl group of type $B_{k-1}$ and thus is symmetric in $\cup_{j=1}^{k-1}\{z_j\}$ as well as invariant under $z_j \leftrightarrow 1/z_j$, $\forall 1\leq j\leq k-1$. Similarly, we also have that $X^{(k)}_L\in\mathbb{C}[z_{k+1}^{\pm},\ldots,z_{L}^{\pm}]^{W_{\rm B}}$. Likewise we can show that $Y^{(k)}_L\in\mathbb{C}[z_1^{\pm},\ldots,z_{k-1}^{\pm}]^{W_{\rm B}} \cap \mathbb{C}[z_{k}^{\pm},\ldots,z_{L}^{\pm}]^{W_{\rm B}}$. There exist further symmetries, but, as we will show now, these require more work to prove.

\subsection{Symmetries of $Y_L$}

We will show that $Y_L^{(k)}$ in fact is independent of $k$, and is symmetric in the combined set of inhomogeneities, $Y^{(k)}_L\in\mathbb{C}[z_1^{\pm},\ldots,z_{L}^{\pm}]^{W_{\rm B}}$.

Consider an infinite lattice made up of double-row transfer matrices. Let $P$ be the set of all paths $p$ which start from the left boundary and end at the right. Since $Y_L^{(k)}$ depends only on the spectral parameter of one double-row transfer matrix, we can assume that the internal spectral parameters of all transfer matrices are equal. Let us call this common value $w$. Therefore a path is independent of its vertical position, and we shall abuse notation to also denote by $p$ the equivalence class of all vertical translates of a particular path $p$. The weight $\omega_p$ of each path is the product of the Boltzmann weights associated to the local path orientations from each $R$ and $K$ matrix involved in fixing $p$. 

The set $P$ can be split into two mutually exclusive subsets: $P_{\rm t}$, the set of paths which start in the top half of the $K$ matrix on the left, and $P_{\rm b}$, the set of paths which start in the bottom half. These sets have a $1$--$1$ correspondence: each path $p\in P_{\rm t}$ can be transformed into a path $\tilde p\in P_{\rm b}$ by a vertical flip. The weights of the two paths are related by $\omega_p(z_1,\ldots,z_L)=\omega_{\tilde p}(z_1^{-1},\ldots,z_L^{-1})$.

Consider now $Y^{(k)}, k\in\{1,\ldots,L+1\}$, and a path $p\in P_{\rm b}$.
Because we must consider all of the possible vertical positions of $p$, it may have multiple contributions to $Y$. The number of contributions $m_p$ depends on the number of times the path crosses back and forth over the vertical line on which the markers of $\widehat Y$ are placed. It can be seen that when $k$ is odd, all paths in $P_{\rm b}$ contribute positively, because each path enters $\widehat Y^{(k)}$ from the left. They all contribute negatively (i.e. enter from the right) when $k$ is even. The opposite is true for the paths in $P_{\rm t}$. 

\begin{figure}[h]
\begin{center}
\begin{picture}(170,90)
\put(0,20){\includegraphics[height=70pt]{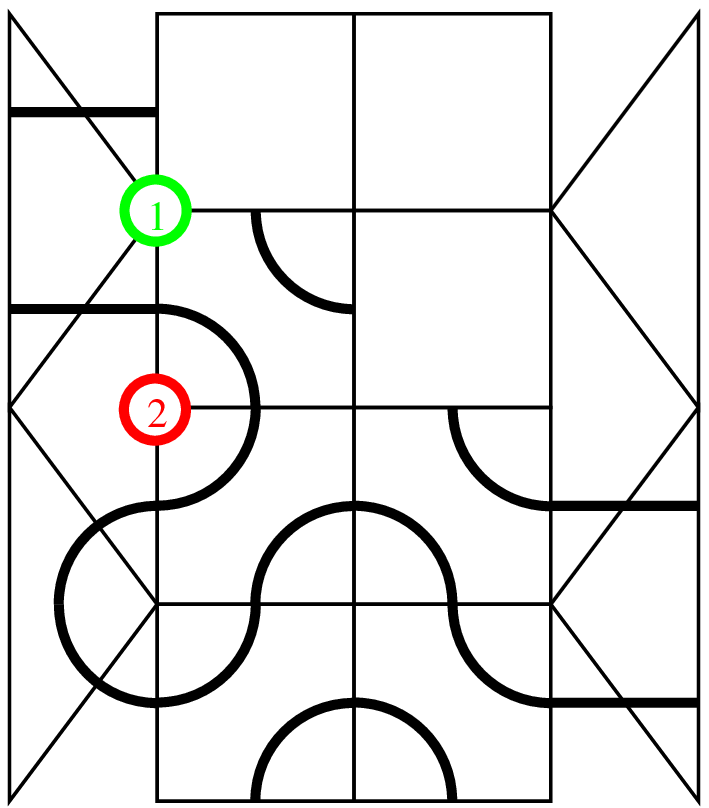}}
\put(0,0){a) path in $P_{\rm b}$}
\put(100,20){\includegraphics[height=70pt]{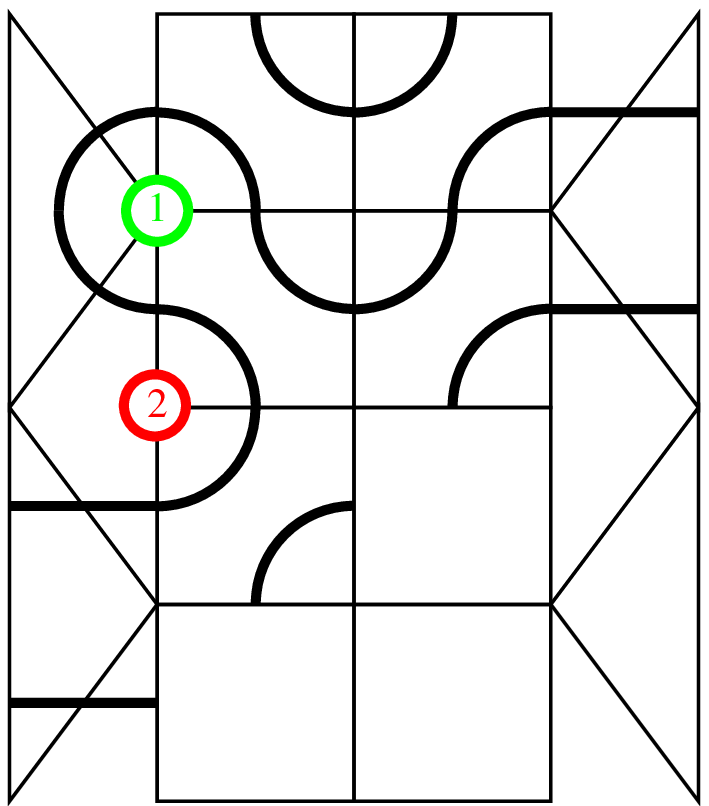}}
\put(100,0){b) path in $P_{\rm t}$}
\end{picture}
\end{center}
\caption{Paths contributing to $Y^{(1)}_2$.}
\label{paths}
\end{figure}

For example, for $k=1$, the path $p_1\in P_{\rm b}$ in Figure~\ref{paths}a) has one vertical translate which also winds through the two markers, hence its multiplicity $m_{p_1}=2$. The sister path $\tilde p_1$, depicted in Figure~\ref{paths}b), is obtained by flipping the complete configuration (but not the markers), and has $m_{\tilde{p_1}}=1$.

Since a path must cross position $k$ from left to right exactly once more than it crosses from right to left, there is always one more path contributing positively than negatively. The consequence of this is that, for $p\in P_{\rm b}$ and its sister path $\tilde p\in P_{\rm t}$, it holds that $m_{\tilde p}=m_p-1$ when $k$ is odd and $m_{\tilde p}=m_p+1$ when $k$ is even. Consequently, $Y$ is given by
\begin{align}
Y^{(k)}_L(z_1,\ldots,z_L)&=(-1)^{k} \left(\sum_{\tilde p\in P_t} m_{\tilde p}\ \omega_{\tilde p}(z_1,\ldots,z_L)-\sum_{p\in P_b}m_p\ \omega_p(z_1,\ldots,z_L) \right)\non\\
&=(-1)^k\sum_{p\in P_b} \Big((m_p+(-1)^k)\omega_p(z_1^{-1},\ldots,z_L^{-1})-m_p\ \omega_p(z_1,\ldots,z_L)\Big)\ .
\end{align}
As we showed in the previous section, $Y^{(k)}_L$ is invariant under $z_i\rightarrow 1/z_i$, for $i<k$ as well as for $i\geq k$, so consider the construction
\begin{align}
Y^{(k)}_L(z_1,\ldots,z_L) &= \frac12\Big(Y^{(k)}(z_1,\ldots,z_L)+Y^{(k)}(z_1^{-1},\ldots,z_L^{-1})\Big)\non\\
&=\frac12(-1)^{k}\sum_{p\in P_b}\Big((m_p+(-1)^k)\omega_p(z_1^{-1},\ldots,z_L^{-1})-m_p\ \omega_p(z_1,\ldots,z_L) \non\\
&\qquad+(m_p+(-1)^k)\omega_p(z_1,\ldots,z_L)-m_p\ \omega_p(z_1^{-1},\ldots,z_L^{-1}) \Big)\non\\
&= \frac12 \sum_{p\in P_b}\Big(\omega_p(z_1^{-1},\ldots,z_L^{-1})+\omega_p(z_1,\ldots,z_L)\Big).
\end{align}
It thus follows that each translational equivalence class of paths $p$ in $P_{\rm b}$ has only one contribution to $Y$, which is equal to the average of the weights of $p$ and $\tilde p$. It is also clear from the above that $Y^{(k)}$ is in fact independent of $k$. Since $Y_L$ is independent of its horizontal position, there is no longer a restriction on the symmetries, so it is symmetric in $z_i$ and invariant under $z_j\rightarrow 1/z_j$, $\forall i,j$.

Furthermore, as a function of the spectral parameter $w$, it now follows from \eqref{Ttranspose} that $Y_L$ is symmetric under $w\rightarrow q/w$,
\be
\label{Ysym}
Y_L(w;\zeta_1,\zeta_2;\ldots) = Y_L(qw^{-1};\zeta_2,\zeta_1;\ldots).
\ee
\subsection{Symmetries of $X^{(k)}_L$}

In Appendix~\ref{ap:YtoX}, the following relationship is proven:
\be
X^{(k)}_L(z_1,\ldots,z_L)=Y_L(w;z_1,\ldots,z_L)|_{w=z_k}.
\label{recur1}
\ee
Based on the symmetries of $Y_L$, it is clear from \eqref{recur1} that $X^{(k)}_L\in\mathbb{C}[z_1^{\pm},\ldots,\hat{z}_k,\ldots,z_{L}^{\pm}]^{W_{\rm B}}$. It also follows that $X^{(k)}=X^{(j)}|_{z_k\leftrightarrow z_j}$.

In addition, if a path $p\in P_b$ has a contribution of $\omega_p(z_1,\ldots,z_L)$ to $X^{(k)}_L$, then its sister path has the contribution $-\omega_p(z_1^{-1},\ldots,z_L^{-1})$. As a result, $X^{(k)}_L$ is antisymmetric when all $z_i$ are sent to $1/z_i$. From the already known symmetries, this antisymmetry must be attributed to $z_k\rightarrow 1/z_k$,
\be
\label{antisym}
X^{(k)}_L(\ldots,z_k,\ldots) = -X^{(k)}_L(\ldots,z_k^{-1},\ldots).
\ee

Finally, note the additivity property around an elementary plaquette,
\be
\label{additivity}
\widehat X^{(k)({\rm bot})}-\widehat Y^{(k+1)}-\widehat X^{(k)({\rm mid})}+\widehat Y^{(k)}=0.
\ee
The invariance of $\widehat Y$ implies that the operator $\widehat X^{({\rm mid})}$, placed in the middle of the transfer matrix, has the same expectation value as an operator $\widehat X$ placed on the bottom.
%
\section{Recursion relations}
In this section, we will list relations that are satisfied by the expectation values $X_L^{(k)}$ and $Y_L$, with a view to show that there are enough such relations, given the degree, to determine these polynomials. Let us first recall the recursions satisfied by the transfer matrix and the normalisation \cite{GierPS}.
%
\subsection{Recursion for the transfer matrix}
Let $\varphi_i$ denote the map that sends site $j$ to $j+2$ for $j\geq i$ in a link pattern, and then inserts a link from site $i$ to $i+1$, thus creating a link pattern of size two greater. For example,
\be
\varphi_3:\quad )(()((\quad \mapsto\quad )({\ir ()}()((.
\ee
In \cite{GierPS} it was shown that for $q=\e^{2\pi\ii/3}$
\be
T_L(w;z_{i+1}=qz_i)\circ\varphi_i=\varphi_i\circ T_{L-2}(w;\hat z_i,\hat z_{i+1}).
\label{Tmatrecur}
\ee
The argument for deriving \eqref{Tmatrecur} is as follows. From the $q$KZ equation \eqref{qKZ} it follows that setting $z_{i+1}=qz_{i}$ in the eigenvector $\ket{\Psi}$ immediately causes many of its entries to vanish. The only nonzero entries are the ones associated to a link pattern with a small link from position $i$ to $i+1$. Let us denote a state corresponding to such a link pattern as $\varphi_i\ket{\alpha_{L-2}}$, where $\alpha_{L-2}$ is a generic link pattern for size $L-2$. We thus consider local configurations of the form
\be
T_L\circ\varphi_i\ \Big|_{z_{i+1}=qz_i}=\quad\raisebox{-40pt}{\includegraphics[height=100pt]{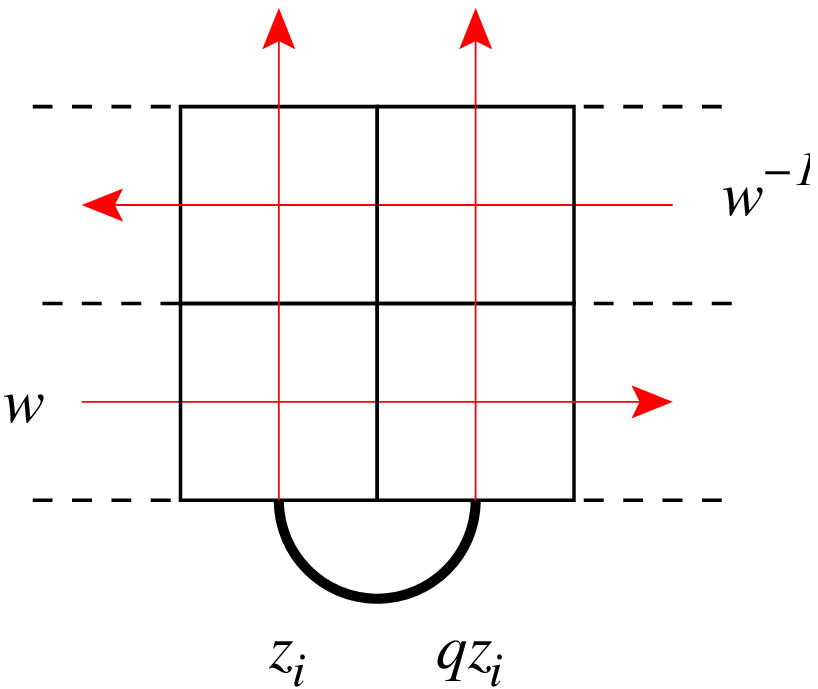}}.
\ee
As shown in Appendix A of \cite{GierPS}, the sixteen possible configurations in the above picture can be grouped by their connectivities, and the weight of each connectivity is then the sum of weights of the orientations in the group. In this way there are five groups of three terms each, with one left over (two of these groups actually have the same connectivity, but it is simpler to consider them separately). The weight associated with each group of three has a factor of the following form:
\be
\frac{[q/qu]}{[q^2u]}\frac{[q/u]}{[qu]}+\frac{[1/qu]}{[q^2u]}\frac{[1/u]}{[qu]}-(q+q^{-1})\frac{[q/qu]}{[q^2u]}\frac{[1/u]}{[qu]}=0,
\ee
for some $u$. Therefore, out of the all the connectivities, only one has nonzero weight:
\be
T_L\circ\varphi_i\ \Big|_{z_{i+1}=qz_i}=\quad\raisebox{-40pt}{\includegraphics[height=90pt]{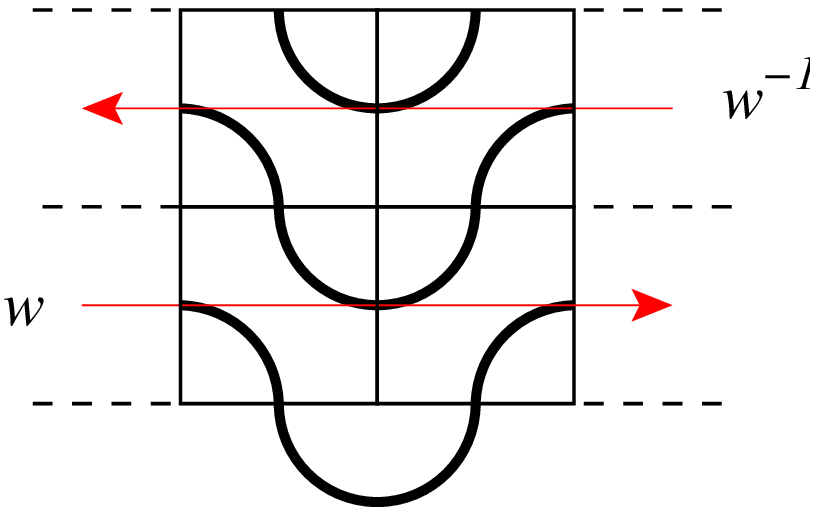}}.
\ee
This weight, when $q=\e^{2\pi\ii/3}$, is equal to $1$, leading to
\begin{align}
\label{eq:bulkrecur}
T_L\circ \varphi_i\ \Big|_{z_{i+1}=qz_i}
=\varphi_i \circ T_{L-2}(\hat z_i,\hat z_{i+1}).
\end{align}

Likewise one can prove that at the boundary, see Appendix \ref{ap:boundrecur}, the transfer matrix at $q=\e^{2\pi\ii/3}$ satisfies
\begin{align}
T_L(w;\zeta_1,\zeta_2;z_1=q\zeta_1,\ldots,z_L) \circ \varphi_0 &= \varphi_0 \circ T_{L-1}(w;q\zeta_1,\zeta_2,z_2,\ldots,z_L),
\label{eq:boundrecur1}\\
T_L(w;\zeta_1,\zeta_2,z_1,\ldots,z_L=q^{-1}\zeta_2) \circ \varphi_L &= \varphi_L \circ T_{L-1}(w;\zeta_1,q^{-1}\zeta_2;z_1,\ldots,z_{L-1}),
\label{eq:boundrecurL}
\end{align}
where $\varphi_0$ is the map that sends site $j$ to site $j+1$, and inserts a closing parenthesis on the first site, and $\varphi_L$ is the map that inserts an opening parenthesis after the last site.

Now, acting with both sides of \eqref{eq:bulkrecur} on the vector $\ket{\Psi(\hat z_i,\hat z_{i+1})}_{L-2}$, we get
\be
T_L(w;z_{i+1}=qz_i)\ \Bigl(\varphi_i \ket{\Psi(\hat z_i,\hat z_{i+1})}_{L-2}\Bigr) = \varphi_i \ket{\Psi(\hat z_i,\hat z_{i+1})}_{L-2}\ ,
\ee
which, by uniqueness of the eigenvector $\ket\Psi_L$, implies that 
\be
\lb{eq:psiprop}\ket{\Psi(z_{i+1}=qz_i)}_L= p(z_i;\ldots,\hat{z}_i,\hat{z}_{i+1}, \ldots)\ \varphi_i \ket{\Psi(\hat z_i,\hat z_{i+1})}_{L-2},
\ee
where $p(z_i;\ldots,\hat{z}_i,\hat{z}_{i+1}, \ldots)$ is a proportionality factor which can be determined \cite{GierPS}. We find likewise proportionality results from the boundary recursions \eqref{eq:boundrecur1} and \eqref{eq:boundrecurL}.
%

\subsection{Normalisation}
Recall that the groundstate eigenvector can be written as
\be
\ket{\Psi} = \sum_{\alpha} \psi_{\alpha} \ket{\alpha}.
\ee
With the explicit result of \cite{GierPS} for the proportionality factor $p(z_i;\ldots,\hat{z}_i,\hat z_{i+1},\ldots)$ the recursion \eqref{eq:psiprop} becomes
\be
\psi_{\varphi_i\circ\alpha}\Big|_{z_{i+1}=qz_i}= k(z_i,\zeta_1)^2k(z_i,\zeta_2)^2\mathop{\prod_{\scriptscriptstyle j=1}}_{\scriptscriptstyle j\neq i}^L k(z_i,z_j)^4\ \psi_{\alpha}(\hat{z}_{i},\hat{z}_{i+1}).
\ee
This recursion, along with the boundary recursions,
\begin{align}
\psi_{\varphi_0\circ\alpha}(\zeta_1)\Big|_{z_{1}=q\zeta_1}&= -k(\zeta_1,\zeta_2)\prod_{j=2}^L k(\zeta_1,z_j)^2\
\psi_{\alpha}(q\zeta_1;\hat{z}_1),\\
\psi_{\varphi_L\circ\alpha}(\zeta_2)\Big|_{z_{L}=q^{-1}\zeta_2}&= -k(\zeta_2^{-1},\zeta_1)\prod_{j=1}^{L-1} k(\zeta_2^{-1},z_j)^2\
\psi_{\alpha}(q^{-1}\zeta_2;\hat{z}_L),
\end{align}
allows us to obtain an expression for the overall normalization $Z_L$ of the vector $\ket\Psi$,
\begin{align}
Z_L(\zeta_1,\zeta_2;z_1,\ldots,z_L)&=\sum_\alpha \psi_\alpha(\zeta_1,\zeta_2;z_1,\ldots,z_L)\non\\
&=\tau_L(z_1,\ldots,z_L)\tau_{L+1}(\zeta_1,z_1,\ldots,z_L)\tau_{L+1}(z_1,\ldots,z_L,\zeta_2)\non\\
&\qquad\times \tau_{L+2}(\zeta_1,z_1,\ldots,z_L,\zeta_2),
\end{align}
with $\tau_L$ as defined in \eqref{Sdef}. This result is based partly on the recursion satisfied by $\tau_L$,
\be
\label{eq:srecur}\tau_L(z_1,\ldots,z_L)\Big|_{z_i=qz_j}=(-1)^L \mathop{\prod_{\scriptscriptstyle l=1}}_{\scriptscriptstyle l\neq i,j}^L k(z_j,z_l)\ \tau_{L-2}(z_1,\ldots,\hat z_i,\hat z_j,\ldots,z_L).
\ee
We will determine $Y_L$ and $X^{(k)}_L$ in a similar way.

\subsection{Recursion relations for $Y_L$}
Placing the markers in $\widehat{Y}_L$ sufficiently far away from the region we look at, we can show that exactly the same recursions satisfied by the transfer matrix in \eqref{eq:bulkrecur}, \eqref{eq:boundrecur1} and \eqref{eq:boundrecurL} are also satisfied by $\widehat{Y}_L$. We then find formulas similar to \eqref{eq:psiprop} for $Y_L$. The symmetries of $Y_L$ mean that these recursions can be generalised, for instance the bulk recursions are true with $z_i = q z_j$, not just $z_i = q z_{i+1}$. Since the proportionality factors in the numerator and denominator of the expectation value $Y_L$ cancel each other, we find the following $4L-4$ bulk recursions for $Y_L$
\be
\label{Yrecur_gen}
Y_L\Big|_{z_i=q^{\pm1}z_j^{\pm1}}=Y_{L-2}(\ldots,\hat z_i,\ldots,\hat z_j,\ldots)\qquad j\neq i,
\ee
and for the boundaries, there are $4$ recursions,
\be
\label{Yrecur_bound}
\begin{array}{rl}
Y_L(\zeta_1;\ldots,z_i,\ldots)\Big|_{z_i=(q\zeta_1)^{\pm1}}&=Y_{L-1}((q\zeta_1)^{\pm1};\ldots,\hat z_i,\ldots)\\
Y_L(\zeta_2;\ldots,z_i,\ldots)\Big|_{z_i=(q^{-1}\zeta_2)^{\pm1}}&=Y_{L-1}((q^{-1}\zeta_2)^{\pm1};\ldots,\hat z_i,\ldots).
\end{array}
\ee
A description of the derivation of the boundary recursions can be found in Appendix \ref{ap:boundrecur}.


From Appendix~\ref{ap:YtoX}, the anti-symmetry \eqref{antisym} of $X^{(k)}_L$ and the symmetry \eqref{Ysym}, we also have the relations
\be
\begin{array}{rl}
Y_L(w;\zeta_1,\zeta_2;z_1,\ldots,z_L)\Big|_{w=z_i^{\pm1}} = \pm X^{(i)}_L(\zeta_1,\zeta_2;z_1,\ldots,z_L),\\
Y_L(w;\zeta_1,\zeta_2;z_1,\ldots,z_L)\Big|_{w=q z_i^{\pm1}} = \mp X^{(i)}_L(\zeta_2,\zeta_1;z_1,\ldots,z_L).
\end{array}
\label{eq:YtoX}
\ee
Viewing \eqref{eq:YtoX} as specialisations for $z_i$, and assuming we know $X_L$, $Y_{L-1}$ and $Y_{L-2}$, we thus find the value of $Y_L$ in $4L+4$ values of $z_i$ .

\subsection{Relations for $X^{(k)}_L$}

The expectation value $X^{(k)}$ satisfies many similar recursion relations, leading to expressions for $X^{(k)}$ at specified values of $z_i$ with $i\neq k$. Without loss of generality we take $i>k$ and obtain $4L-8$ recursions for the bulk,
\be
\label{Xrecur_gen}
\begin{array}{rl}
X^{(k)}_L\Big|_{z_i=q^{\pm1}z_j^{\pm1}} &= X^{(k)}_{L-2}(\ldots,\hat z_i,\ldots,\hat z_j,\ldots)\qquad j>k\\
X^{(k)}_L\Big|_{z_i=q^{\pm1}z_j^{\pm1}} &= X^{(k-1)}_{L-2}(\ldots,\hat z_j,\ldots,\hat z_i,\ldots)\qquad j<k\\
\end{array}
\ee
and $4$ recursions for the boundaries,
\be
\label{Xrecur_bound}
\begin{array}{rl}
X^{(k)}_L(\zeta_1;\ldots,z_i,\ldots)\Big|_{z_i=(q/\zeta_1)^{\pm1}}&=X^{(k)}_{L-1}(q^{\pm1}\zeta_1^{\pm1};\ldots,\hat z_i,\ldots)\\
X^{(k)}_L(\zeta_2;\ldots,z_i,\ldots)\Big|_{z_i=(q\zeta_2)^{\pm1}}&=X^{(k)}_{L-1}(q^{\pm1}\zeta_2^{\pm1};\ldots,\hat z_i,\ldots)
\end{array}
\ee
Furthermore, Appendix~\ref{ap:YtoX} and the anti-symmetry \eqref{antisym} of $X^{(k)}_L$ under $z_k \rightarrow 1/z_k$ imply four more recursions,
\be
\label{eq:XtoY}
\begin{array}{rl}
X^{(k)}_L\Big|_{z_i=(z_k/q)^{\pm1}} &=Y_{L-2}(z_k;\hat z_i,\hat z_k)\qquad k\neq i \\
X^{(k)}_L\Big|_{z_i=(qz_k)^{\pm1}} &=-Y_{L-2}(z_k;\hat z_i,\hat z_k)\qquad k\neq i
\end{array}\ .
\ee
Therefore, assuming we know $Y_{L-2}$, $X^{(k)}_{L-2}$ and $X^{(k)}_{L-1}$, we know the value of $X^{(k)}_L$, viewed as a polynomial in $z_i$, in $4L$ points.

%
\section{Solution}
\subsection{Degree}
\label{degree}
We recall definition \eqref{Expectation} of the current expectation value
\be
F_L^{(x_1,x_2)}=\frac{1}{Z_L^2} \bra{\Psi^*}\ \widehat{F}_L^{(x_1,x_2)}\ \ket{\Psi}.
\ee
We have observed for small systems, that the numerator contains a factor $Z_L$, cancelling one of such factors in the denominator. We assume that this factorisation occurs for all system sizes, and hence that $F_L$ is of the form $F_L^{(x_1,x_2)}=P/Z_L$, where $P$ is a polynomial. Unfortunately we have not been able to find a proof for this factorisation.

For both expectation values $X_L$ and $Y_L$, the degree width of the numerator must equal the degree width of the denominator. The degree width of $Z_L$ is $4L-2$ in each variable $z_i$. By the assumption above, this means for $X^{(k)}_L$ that the numerator must also be of degree width $4L-2$ in each variable. For $Y_L$, we must also take into account the action of the transfer matrix. The numerator and denominator of each term in the transfer matrix both have degree width $4$ in each variable. We can factor the denominator out, so the degree width of the numerator of $Y_L$ is now $4L+2$. Thus we have enough recursion relations to fix the numerators of both expectation values.

\subsection{Proof of the main result}
Under the hypothesis outlined in the previous section, we now prove the form of the expectation values as given in equation \eqref{Xval} and \eqref{Yval},
\begin{align}
X^{(i)}_L &= (-)^L\ii\frac{\sqrt{3}}{2}\ z_i \frac{\partial}{\partial z_i}\ u_L(\zeta_1,\zeta_2;z_1,\ldots,z_L),
\label{Xform}\\
Y_L &= (-)^L\ii\frac{\sqrt{3}}{2}\left.\ w \frac{\partial}{\partial w}\ u_{L+2}(\zeta_1,\zeta_2;z_1,\ldots,z_L,qv^{-1},w)\right|_{v=w},
\label{Yform}
\end{align}
with
\be
\label{eq:u}
u_L(\zeta_1,\zeta_2;z_1,\ldots,z_L)=\log \left[ \frac{\tau_{L+1}(\zeta_1,z_1,\ldots,z_L) \tau_{L+1}(\zeta_2,z_1,\ldots,z_L)} {\tau_{L}(z_1,\ldots,z_L) \tau_{L+2}(\zeta_1,\zeta_2,z_1,\ldots,z_L)} \right].
\ee
The function $u$ has a list of simple recursions, based on \eqref{eq:srecur}:
\begin{align}
u_L(\zeta_1,\zeta_2;z_1,\ldots,z_{L-2},z,qz) &= u_{L-2}(\zeta_1,\zeta_2;z_1,\ldots,z_{L-2}),\nonumber\\
u_L(\zeta_1,\zeta_2;z_1=q\zeta_1,z_2,\ldots,z_{L}) &= -u_{L-1}(q\zeta_1,\zeta_2;z_2,\ldots,z_{L})-\log\left[-k(\zeta_1,\zeta_2 )\right ],\\
u_L(\zeta_1,\zeta_2;z_1,\ldots,z_{L-1},z_L=q^{-1}\zeta_2) &= -u_{L-1}(\zeta_1,q^{-1}\zeta_2;z_1,\ldots,z_{L-1})-\log\left[-k(q^{-1}\zeta_2,\zeta_1 )\right ]. \nonumber
\end{align}
The first of these is easily obtained by using the recursion \eqref{eq:srecur} on each of the $\tau$-functions in \eqref{eq:u}, and noting that the proportionality factors on the top and bottom cancel exactly. The other two recursions work in a similar way, but result in the reciprocal of the needed ratio of $\tau$-functions, giving a minus sign in front of the logarithm. In fact, the proportionality factors do not cancel completely, but produce an extra term which is dependent only on $\zeta_1$ and $\zeta_2$, and thus disappears when the derivative is taken in \eqref{Xform} and \eqref{Yform}.

These recursions immediately imply the recursions for $Y_L$ and $X_L^{(i)}$ which do not involve the marked parameter; that is, the $4L$ recursions in \eqref{Yrecur_gen} and \eqref{Yrecur_bound} for $Y_L$, as well as the $4L-4$ recursions in \eqref{Xrecur_gen} and \eqref{Xrecur_bound} for $X_L$. The remaining recursions \eqref{eq:YtoX} and \eqref{eq:XtoY}, which respectively complete the degree arguments for $Y_L$ and $X_L^{(i)}$, are more complicated but are still satisfied by the above forms. These are proved in the next section.

Thus, to finish the proof we simply need to verify the formul\ae\ \eqref{Xform} and \eqref{Yform} for small values of $L$ as initial conditions for the recursions. We have checked by direct calculation that \eqref{Xform} and \eqref{Yform} indeed hold for $L=2$ and $L=3$.

%
\subsection{Proof of recursions \eqref{eq:YtoX} and \eqref{eq:XtoY}}
We will first prove that $Y_L(w)$ as given by \eqref{Yform} obeys the symmetry \eqref{Ysym}, i.e.
\be
Y_L(w;\ldots) = Y_L(q/w;\ldots).
\ee
This follows from
\begin{align}
-w \frac{\partial}{\partial w}\ u_{L+2}(qv^{-1},qw^{-1})\Big|_{v=q/w} &= -v \frac{\partial}{\partial v}\ u_{L+2}(w,qv^{-1})\Big|_{v=w} \non\\
&= v \frac{\partial}{\partial v}\ u_{L+2}(w,v)\Big|_{v=q/w} \non\\
&= w \frac{\partial}{\partial w}\ u_{L+2}(w,qv^{-1})\Big|_{v=w},
\end{align}
where we have used that $u_L(z_1,\ldots,z_L)$ is symmetric in $\{z_1^\pm,\ldots,z_L^\pm\}$.

Now we prove relation \eqref{eq:YtoX} for the case where $z_i=w$. The other cases follow by the anti-symmetry \eqref{antisym} of $X^{(i)}_L$ and the symmetry \eqref{Ysym}, proved above. We have
\begin{align}
Y_L\Big|_{w=z_i}
&=(-)^L\ii\frac{\sqrt{3}}{2}\left.\ w \frac{\partial}{\partial w}\ u_{L+2}(qv^{-1},w)\right|_{\begin{subarray}{l}v=z_i\\w=z_i\end{subarray}}\non\\
&=(-)^L\ii\frac{\sqrt{3}}{2}\left.\ w \frac{\partial}{\partial w}\ u_{L}(\hat z_i,w)\right|_{w=z_i}\non\\
&=(-)^L\ii\frac{\sqrt{3}}{2}\ z_i\ \frac{\partial}{\partial z_i}\ u_{L}\non\\
&= X^{(i)}_L(\zeta_1,\zeta_2;z_1,\ldots,z_L).
\end{align}

For relation \eqref{eq:XtoY}, we will prove the case where $z_i=z_k/q$, and the rest follow again from (anti-)symmetry. We have
\begin{align}
X^{(k)}_L\Big|_{z_i=z_k/q}&= (-)^L\ii\frac{\sqrt{3}}{2}\left.\ z_{k} \frac{\partial}{\partial z_{k}}\ u_{L}(z_i,z_k)\right|_{z_i=z_k/q}\non\\
&= (-)^L\ii\frac{\sqrt{3}}{2}\left.\ z_k \frac{\partial}{\partial z_k}\ u_{L}(vq^{-1},z_k)\right|_{v=z_k}\non\\
&= Y_{L-2}(z_k;\hat z_i,\hat z_k).
\end{align}

\section{Conclusion}
In the context of percolation, we have analytically computed an observable, which in a generalised Chalker-Coddington model is the spin current at a quantum Hall transition. It takes a universal value, of which the continuum limit was calculated by Cardy \cite{cardy00}. This observable also falls within a class of parafermionic correlation functions which can be shown to be discretely holomorphic, and hence are precursors to analytic correlators in the conformally invariant continuum limit. Our main results \eqref{Xval} and \eqref{Yval} are expressed in terms of a Toda lattice wave function, which hints at an interesting link between the quantum integrable completely packed O($n=1$) model and classical integrable models. We stress furthermore that our result is exact for systems of finite width, and not asymptotic, which is unusual for systems which are not free fermionic. Analogous results have been obtained for cylindrical boundary conditions for both site- and bond-percolation \cite{Nienhuis10}.

\section*{Acknowledgment}
It is a pleasure to thank Luigi Cantini and Omar Foda for discussions about the Toda lattice. JdG and AP would like to thank the Australian Research Council for financial support, and AP would like to thank the hospitality of the Henri Poincar\'e Institute in Paris and The Rudolf Peierls Institute for Theoretical Physics in Oxford where part of this work was completed. BN would like to thank the hospitality and financial support provided by the ARC Centre of Excellence MASCOS and the Department of Mathematics and Statistics at the University of Melbourne, as well as the financial support from the organisation FOM (which is part of NWO). 
\appendix

\section{Relations between $Y$ and $X$}
\label{ap:YtoX}
Recursion relations between $\widehat{X}$ and $\widehat{Y}$ are obtained by looking at the way the markers can move around in configurations with fixed local orientations of paths. We first determine the recursion that expresses $\widehat{X}_L$ in terms of $\widehat{Y}_{L-2}$,

\begin{align}
\widehat{X}^{(i+1)}_L\Big|_{\begin{subarray}{l} z_{i+1}=w\\z_i = wq^{-1} \end{subarray}}\circ\varphi_i &=\raisebox{-45pt}{\includegraphics[height=100pt]{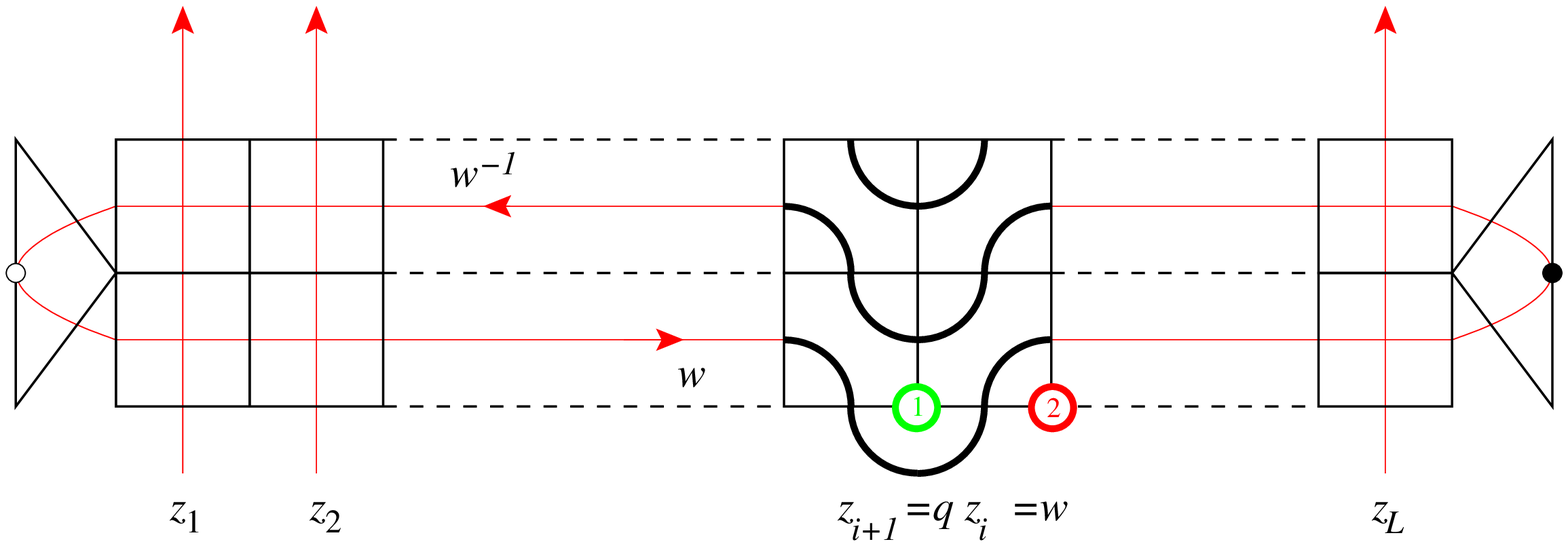}}\nonumber\\
&=\raisebox{-45pt}{\includegraphics[height=100pt]{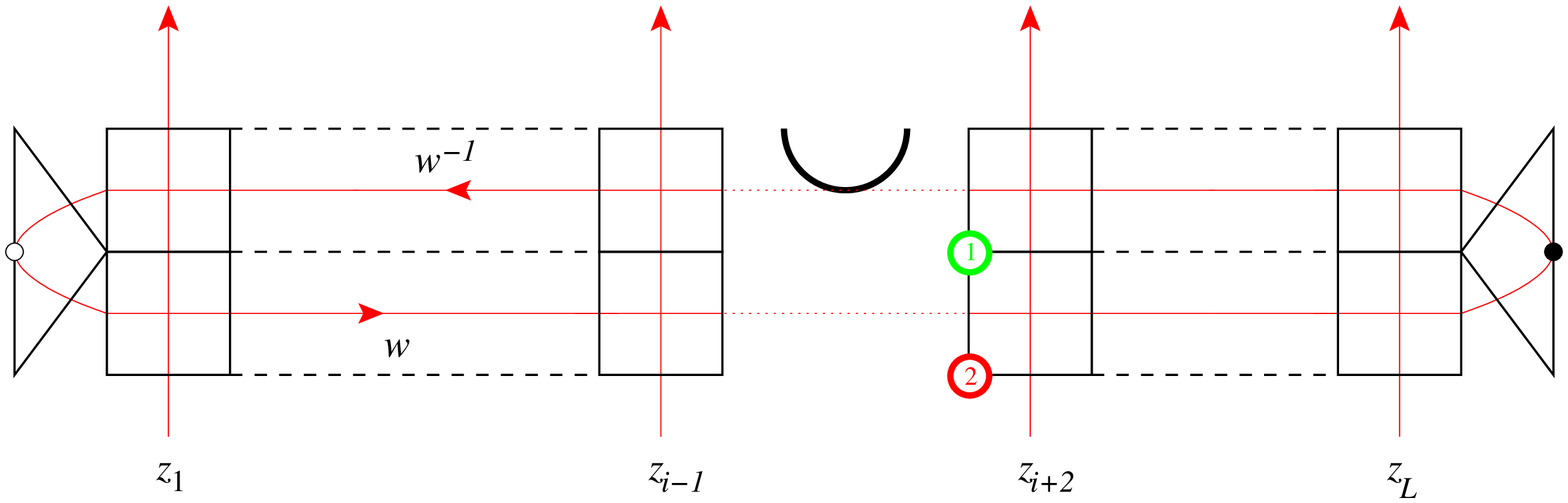}} \nonumber\\
&=\varphi_{i}\circ\widehat{Y}_{L-2}(w;\ldots,\hat{z}_{i},\hat{z}_{i+1},\ldots).
\end{align}

Likewise we can express $\widehat{Y}_L$ in terms of $\widehat{X}_L$. This relation is
\begin{align*}
\widehat{Y}_L|_{w=z_i}&=\raisebox{-45pt}{\includegraphics[height=100pt]{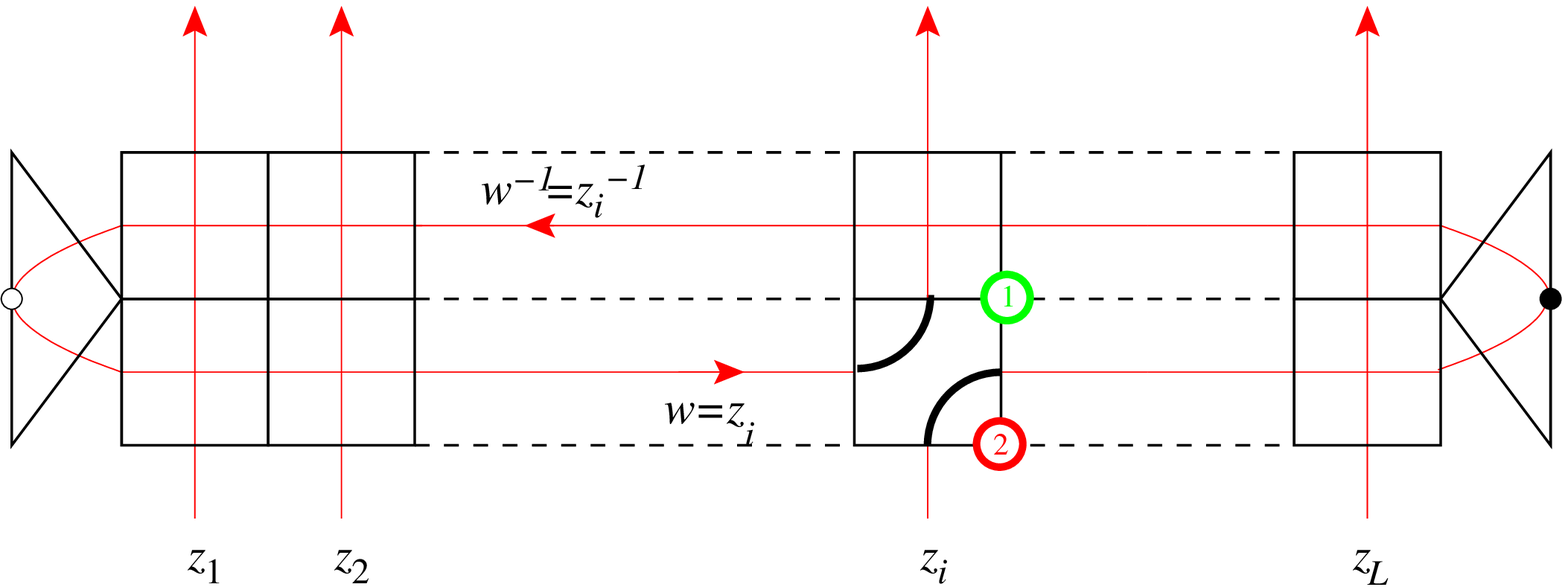}}\\
&=\raisebox{-45pt}{\includegraphics[height=100pt]{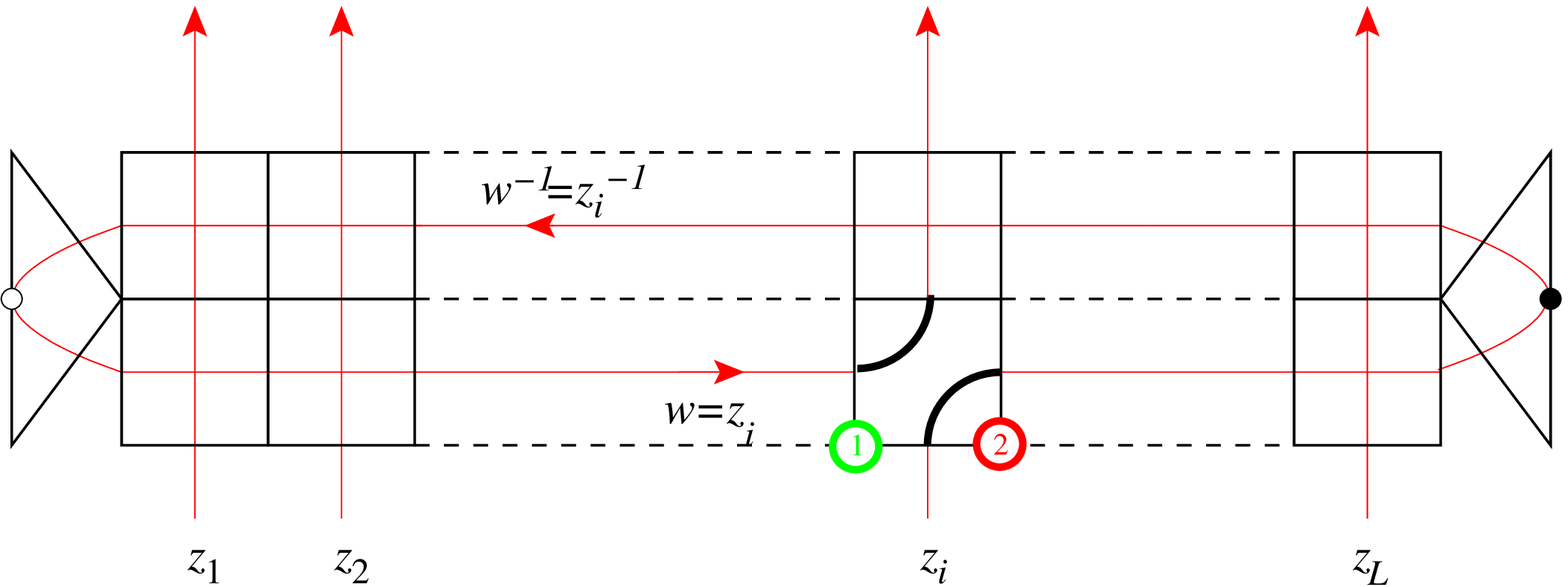}}\\
&=\widehat{X}^{(i)}_L.
\end{align*}
%

\section{The left boundary recursion for the transfer matrix}
\label{ap:boundrecur}
We will describe the recursion of the transfer matrix at the left boundary. The recursion at the right boundary is similar.

Setting $z_1=q\zeta_1$, the entries of the eigenvector are zero unless they correspond to link patterns with a little link from position 1 to the left boundary.
\be
T_L\circ\varphi_0\Big|_{z_1=q\zeta_1}=\quad\raisebox{-40pt}{\includegraphics[height=100pt]{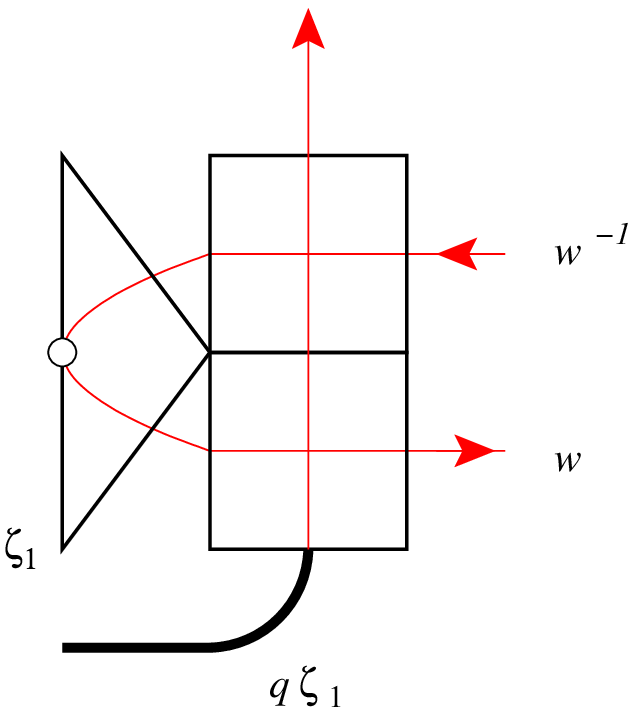}}.
\ee

There are $8$ possible orientations in the above picture, which can be grouped into 3 kinds of connectivities. The following is a summary of the connectivities along with their weights:

\begin{align*}
\includegraphics[height=50pt]{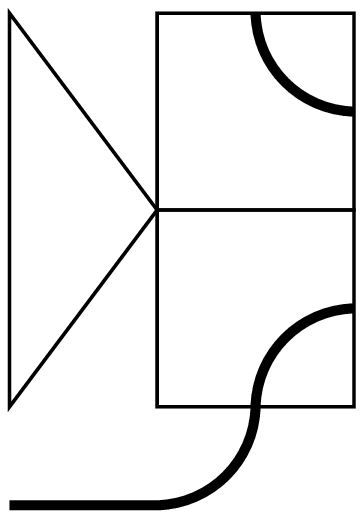}\hspace{2cm}
&\begin{array}{l}
\displaystyle{\frac{[q/q\zeta_1w]}{[q^2\zeta_1w]}\left(\frac{[q^2\zeta_1/w]}{[qw/q\zeta_1]}+\frac{[q\zeta_1/w]}{[qw/q\zeta_1]}\right)\left(\frac{k(q/w,\zeta_1)}{k(w/q,\zeta_1)}+\frac{[1/q][q^2/w^2]}{k(w/q,\zeta_1)}\right)}\vspace{.1cm} \\
\displaystyle{\quad+\frac{[1/q\zeta_1w]}{[q^2\zeta_1w]}\frac{[q^2\zeta_1/w]}{[qw/q\zeta_1]}\frac{k(q/w,\zeta_1)}{k(w/q,\zeta_1)}}\vspace{.1cm} \\
\displaystyle{=\frac{[1/\zeta_1w]}{[q^2\zeta_1w]}+\frac{[1/q\zeta_1w][q^2\zeta_1/w][qw\zeta_1/q][qw/q\zeta_1]}{[q^2\zeta_1w][w/\zeta_1][q^2/w\zeta_1][q^2\zeta_1/w]}}\vspace{.1cm} \\
\displaystyle{=\frac{[1/\zeta_1w]}{[q^2\zeta_1w]}+\frac{[w\zeta_1]}{[q^2\zeta_1w]}=0;}
\end{array}\\
\vspace{.7cm}\\
\includegraphics[height=40pt]{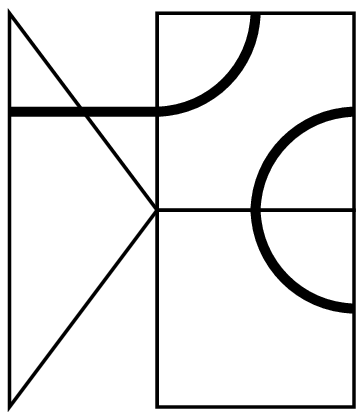}\hspace{2cm}
&\begin{array}{l}
\displaystyle{\frac{[1/q\zeta_1w]}{[q^2\zeta_1w]}\frac{[q\zeta_1/w]}{[qw/q\zeta_1]}\left(\frac{k(q/w,\zeta_1)}{k(w/q,\zeta_1)}+\frac{[1/q][q^2/w^2]}{k(w/q,\zeta_1)}\right)}\vspace{.1cm} \\
\displaystyle{=\frac{[q\zeta_1/w][1/q\zeta_1w]}{[w/\zeta_1][q^2\zeta_1w]}}\vspace{.1cm} \\
\displaystyle{=\frac{k(q/w,q\zeta_1)}{k(w/q,q\zeta_1)};}
\end{array}\\
\vspace{.7cm}\\
\includegraphics[height=50pt]{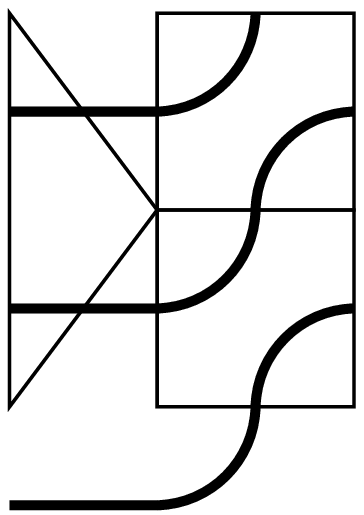}\hspace{2cm}
&\begin{array}{l}
\displaystyle{\frac{[1/q\zeta_1w]}{[q^2\zeta_1w]}\frac{[q^2\zeta_1/w]}{[qw/q\zeta_1]}\frac{[1/q][q^2/w^2]}{k(w/q,\zeta_1)}}\vspace{.1cm} \\
\displaystyle{=\frac{[1/q\zeta_1w][q^2\zeta_1/w][1/q][q^2/w^2]}{[q^2\zeta_1w][w/\zeta_1][q^2\zeta_1/w][q^2/w\zeta_1]}}\vspace{.1cm} \\
\displaystyle{=\frac{[1/q][q^2/w^2]}{[w/\zeta_1][q^2\zeta_1w]}}\vspace{.1cm} \\
\displaystyle{=\frac{[1/q][q^2/w^2]}{k(w/q,q\zeta_1)}.}
\end{array}
\end{align*}
The whole sum can therefore be written as $\varphi_0\circ K_{\rm l}(w,q\zeta_1)$, in a system of size $L-1$. This implies
\be
T_L(\zeta_1;z_1,\ldots)\circ\varphi_0\Big|_{z_1=q\zeta_1}=\varphi_0\circ T_{L-1}(q\zeta_1;z_2,\ldots).
\ee
\newpage
\newcommand\arxiv[1]{\href{http://arxiv.org/pdf/#1}{\texttt{arXiv:#1}}}


\begin{thebibliography}{99}
%
\bibitem{Smir06} S.~Smirnov, \textit{Towards conformal invariance of 2D lattice models}, in Proceedings of the International Congress of Mathematicians (Madrid, August 22-30, 2006), European Mathematical Society, 2006, Volume II, 1421-1451.
%
\bibitem{Smir07} S.~Smirnov, \textit{Conformal invariance in random cluster models. I. Holomorphic fermions in the Ising model}, \arxiv{0708.0039}, to appear in Ann. Math. 172 (2010).
%
\bibitem{RivaCar} V.~Riva and J.~Cardy, \textit{Holomorphic Parafermions in the Potts model and SLE }, J. Stat. Mech. \textbf{0612} (2006), P001, \arxiv{cond-mat/0608496}.
%
\bibitem{IkhCar} Y.~Ikhlef and J.~Cardy, \textit{Discretely Holomorphic Parafermions and Integrable Loop Models}, J. Phys. A \textbf{42} (2009), 102001, \arxiv{0810.5037}.
%
\bibitem{BaxterKW76} R.J.~Baxter, S.B.~Kelland and F.Y.~Wu, \textit{Equivalence of the Potts model or Whitney polynomial with an ice-type model}, J.~Phys.~A: Math.~Gen. \textbf{9} (1976), 397--406.
%
\bibitem{Baxter} R.J.~Baxter, \textit{Exactly solved models in statistical mechanics} (Dover, USA, 2007). Reprint of the 1982 original (Academic Press, London).
%
\bibitem{Gruzberg99} I.A.~Gruzberg, A.W.~W.~Ludwig and N.~Read,
\textit{Exact exponents for the spin quantum Hall transition}, Phys. Rev. Lett. 82 (1999) 4524; \arxiv{cond-mat/9902063}.
%
\bibitem{ChalkC88} J.T.~Chalker and P.D.~Coddington, \textit{Percolation, quantum tunnelling and the integer Hall effect }, J.~Phys.~C \textbf{21} (1988), 2665.
%
\bibitem{Metz99} M.~Metzler, \textit{The Influence of Percolation in the Generalized Chalker-Coddington Model}, J. Phys. Soc. Jpn. \textbf{68} (1999), 144--150
%
\bibitem{Nienhuis10} B.~Nienhuis, J.L.~Jacobsen and P.~Di~Francesco, to be published.
%
\bibitem{JimboM} M.~Jimbo and T.~Miwa, \textit{Solitons and infinite dimensional Lie algebras}, Publ. RIMS \textbf{19} (1983), 943--1001.
%
\bibitem{Skly88} E.K.~Sklyanin, \textit{Boundary conditions for integrable quantum systems}, J.~Phys.~A \textbf{21} (1988), 2375--2389. 
%
\bibitem{GierPS} J.~de Gier, A.~Ponsaing and K.~Shigechi, {\it Exact finite size groundstate of the O($n=1$) loop model with open boundaries}, J. Stat. Mech. (2009), P04010, \arxiv{0901.2961}.
%
\bibitem{RasmusP} P.A.~Pearce, J.~Rasmussen, J.-B.~Zuber, \textit{Logarithmic minimal models}, J.~Stat.~Mech. (2006) P11017, \arxiv{hep-th/0607232}; P.A.~Pearce, J.~Rasmussen, \textit{Polymers, percolation and fusion}, Proceedings of RAQIS'07, Annecy, France (2007).
%
\bibitem{DF05} P.~Di Francesco, \textit{Inhomogeneous loop models with boundaries}, J.~Phys.~A \textbf{38} (2005), 6091--6120, \arxiv{math-ph/0504032}.
%
\bibitem{smirnov} F.A.~Smirnov, \textit{Form Factors in Completely Integrable Models of Quantum Field Theory}, Adv.~Series in Math.~Phys. \textbf{14} (World Scientific, Singapore, 1992).
%
\bibitem{FR} I.B.~Frenkel and N.~Reshetikhin, \textit{Quantum affine algebras and holonomic difference equations}, Commun.~Math.~Phys. \textbf{146} (1992), 1--60.
%
\bibitem{ZJ07} P.~Zinn-Justin, {\it Loop model with mixed boundary conditions, $q$KZ equation and Alternating Sign Matrices}, J.~Stat.~Mech. (2007) P01007, \arxiv{math-ph/0610067}.
%
\bibitem{Cant} L.~Cantini, \textit{qKZ equation and ground state of the O(1) loop model with open boundary conditions}, (2009), \arxiv{0903.5050}.
%
\bibitem{cardy00} J.L.~Cardy, \textit{Linking numbers for self-avoiding walks and percolation: application to the spin quantum Hall transition}, Phys. Rev. Lett. 84 (2000) 3507; \arxiv{cond-mat/9911457}

\end{thebibliography}
\end{document}